\documentclass[twocolumn]{aastex631}
\usepackage{color}
\usepackage{amsmath}
\usepackage{amsfonts}
\usepackage{amsthm}
\usepackage{bm}
\usepackage{booktabs}
\usepackage{natbib}
\usepackage{bm}
\DeclareGraphicsExtensions{.png,.jpg,.pdf,.eps}
\def\St{{\rm{St}} }

\received{\today}

\shorttitle{Length and Velocity scales in turbulent nebula.}
\shortauthors{Sengupta et. al.}

\definecolor{purple}{rgb}{0.7,0.0,0.7}

\definecolor{brown}{rgb}{0.42,0.24,0.07}

\newcommand{\bug}{{\bf u}}
\newcommand{\bup}{{\bf u_p}}

\newcommand{\vk}{v_k}
\newcommand{\tanstress}{{\cal{W}}_{r\phi}}
\newcommand{\alphat}{\Tilde{\alpha}}
\newcommand{\zrms}{\sqrt{\langle z^2\rangle}}
\newcommand{\Ro}{${\rm Ro}$}

\begin{document}


\title{Length and Velocity Scales in Protoplanetary Disk Turbulence.}

\author[0000-0003-0801-3159]{Debanjan Sengupta}
\affiliation{Department of Astronomy, New Mexico State University, PO Box 30001, MSC 4500, Las Cruces, NM 88003-8001, USA; (debanjan@nmsu.edu)}
\affiliation{NASA Postdoctoral Program Fellow, Ames Research Center, NASA; Mail Stop 245-3, Moffett Field, CA 94035, USA}

\author[0000-0003-0553-1436]{Jeffrey N. Cuzzi}
\affiliation{NASA Ames Research Center; Mail Stop 245-3, Moffett Field, CA 94035, USA}

\author[0000-0001-5372-4254]{Orkan M. Umurhan}
\affiliation{SETI Institute, 389 Bernardo Way, Mountain View, CA 94043, USA}
\affiliation{NASA Ames Research Center; Mail Stop 245-3, Moffett Field, CA 94035, USA}
\affiliation{Cornell Center for Astrophysics and Planetary Sciences, Cornell University, Ithaca, NY 14853, USA}

\affiliation{Department of Earth and Planetary Science, University of California Berkeley, Berkeley, CA 94720, USA}

\author[0000-0002-3768-7542]{Wladimir Lyra}
\affiliation{Department of Astronomy, New Mexico State University, PO Box 30001, MSC 4500, Las Cruces, NM 88003-8001, USA; (debanjan@nmsu.edu)}



\begin{abstract}

In the theory of protoplanetary disk turbulence, a widely adopted  \emph{ansatz}, or assumption, is that the  turnover frequency of the largest turbulent eddy, $\Omega_L$, is the local Keplerian frequency $\Omega_K$. In terms of the standard dimensionless Shakura-Sunyaev $\alpha$ parameter that quantifies turbulent viscosity or diffusivity, this assumption leads to characteristic length and velocity scales given respectively by $\sqrt{\alpha}H$ and $\sqrt{\alpha}c$, in which $H$ and $c$ are the local gas scale height and sound speed. However, this assumption is not applicable in cases when turbulence is forced numerically or driven by some natural processes such as Vertical Shear Instability. Here we explore the more general case where $\Omega_L\ge\Omega_K$ and show  that under these conditions, the characteristic length and velocity scales are respectively $\sqrt{\alpha/R'}H$ and $\sqrt{\alpha R'}c$, where $R'\equiv \Omega_L/\Omega_K$ is twice the Rossby number. It follows that $\alpha=\alphat/R'$, where $\sqrt{\alphat} c$ is the root-mean-square average of the turbulent velocities. Properly allowing for this effect naturally explains the reduced particle scale heights produced in shearing box simulations of particles in forced turbulence, and may help with interpreting recent edge-on disk observations; more general implications for observations are also presented. For $R'>1$ the effective particle Stokes numbers are increased, which has implications for particle collision dynamics and growth,  as well as for planetesimal formation.

\end{abstract}

\section{Introduction}\label{sec:introduction}
Protoplanetary disks are increasingly regarded as being moderately turbulent, under a handful of magnetohydrodynamic \citep{balbus_Hawley_1991, Turner_etal_2014, Lesur_etal_2022} or purely hydrodynamic instabilities \citep{Nelson_etal_2013, Lyra_2014, Marcus_etal_2013}. In many applications, such as global models of disk evolution under turbulent viscosity \citep{Estrada_etal_2016, Sengupta_etal_2022}, it is impractical to use more than a simple parameterization of the viscous and diffusive effects of turbulence, and by far the most popular one is the so-called ``$\alpha$-model" \citep{Shakura_Sunyaev_1973, Shakura_etal_1978}. In this closure model, the spatial and temporal evolution of turbulent protoplanetary gas and particle disks is determined by the large-scale effective viscosity and diffusivity of the gas, and the particles in it. For many purposes, it is adequate (and indeed necessary given our limited understanding) to simplify the complex effects of real turbulence into a scalar viscosity $\nu$ and diffusivity $D$. The basic scaling or mixing length approximation for turbulent diffusivity and viscosity is $\left(D, \nu\right) \equiv [L][V_L]$,  where $L$ and $V_L$ are respectively the characteristic length and velocity scales of the turbulence. In this case, $L$  can be identified with the energy injection scale or the energy containing large spatial scale and $V_L$ is the velocity at that scale. \citet{Shakura_Sunyaev_1973}  introduced the widely-used mixing length-like closure model $\nu \equiv \alpha c H$ to incorporate our ignorance about the turbulence into the single parameter $\alpha$, where $c$ is the gas sound speed and $H$ is the gas vertical scale height. In this classic paper\footnote{We note that the first description and application of mixing length theory to disks may be found in \citet{Prendergast_Burbidge_1968}, but an explicit mathematical prescription was not given.}, it was suggested briefly that $L \sim H$ and thus $V_L \sim \alpha c$, indeed with $\alpha$ often taken to be of order unity \citep{Cameron_1978, Lin_Bodenheimer_1982, Weidenschilling_1984, Morfill_1985, Weidenschilling_Cuzzi_1993}, given that early works were more focused on accretion disks around compact objects where trans-sonic turbulence used to be widely believed. Moreover, to study the properties of solid particles (hereafter called simply ``particles") in the disk as they grow by collisional sticking and evolve radially by drift and diffusion \citep[e.g.,][]{Estrada_etal_2016, Estrada_etal_2021, Estrada_etal_2022}, it has become necessary to separate the velocity and length scale properties of the turbulence that influence the particles \citep{Cuzzi_etal_2001}. 

A common assumption in such studies is that the turnover time of the largest, energy-containing, eddy of the nebula turbulence equals the local orbital time \citep{Shakura_Sunyaev_1973}. \citet[Appendix I and page 184]{Shakura_etal_1978} worked through an energy dissipation argument that implied $\Omega_L =\Omega_K$, and a logical chain resulted similar to that given in section \ref{sec:omegaL=omegaK} below (see Appendix \ref{sec:Shakuraetal1978} for more details). The $\Omega_L=\Omega_K$ assumption had been made earlier by \citet{Safronov_1972}, and there has been some numerical support for this assumption under certain conditions \citep{Coleman_etal_1992}. 

As simulations of particle collisions and growth in turbulence have become more detailed, it has also become important to reasonably estimate particle velocities given $V_L$ \citep{Voelk_etal_1980}. \citet{Weidenschilling_1997} adopted the scaling of \citet{Shakura_etal_1978}, and \citet{Cuzzi_etal_2001} independently reproduced the scaling of \citet{Shakura_etal_1978} in the form most often used today, as described below in section \ref{sec:omegaL=omegaK}. \citet{Dubrulle_1992} generalized the discussion to eddy frequencies $\Omega_L > \Omega_K$ and  $\Omega_L < \Omega_K$, but concluded that $\Omega_L \sim \Omega_K$ for self-generated nebula turbulence. \citet{Dubrulle_1992} even proposed a scaling of $L$ relevant to these conditions, but without connecting it to $\alpha$ in any clear way. The idea that large eddy frequencies in turbulence could significantly exceed the orbit frequency has also been discussed in the context of midplane turbulence generated by a settled particle layer \citep{Cuzzi_etal_1993, Sekiya_1998}, but has never been explored using direct numerical simulations.

$\Omega_L=\Omega_K$ seems to be a sensible assumption given that the disk is rotating, and thus any radial or azimuthal velocity deviation will be subsequently dominated by the Coriolis force, leading to a Keplerian turnover time \citep[e.g.][]{Cuzzi_etal_2001}. While reasonable for the magnetorotational instability \citep[MRI,][]{balbus_Hawley_1991}, this result does not appear to be generally valid for turbulence generated by e.g., Vertical Shear Instability, where \citet{Stoll_Kley_2016} measure eddy lifetimes much shorter than Keplerian, of the order of $0.1\Omega_K^{-1}$. \citet{Sengupta_Umurhan_2023} find similarly short eddy lifetimes for the turbulence generated by the settled particle layer at the disk midplane. Incidentally, even for the MRI, \citet{Johansen_et_al_2006} find eddy lifetimes to be shorter than Keplerian in the case of forcing by a sufficiently strong externally imposed vertical magnetic field. 

Taken all in all, these results suggest that the tendency of the Coriolis force to impose a Keplerian turnover time will be in competition with other external forcing agents acting on the system. If these other forcing agents exert greater strength or amplitude, then the Coriolis force becomes less influential, potentially resulting in the turnover time of the largest eddies deviating from Keplerian. It is this case we investigate in this paper.

In this contribution, we  numerically assess the properties of externally forced turbulence under conditions of a rotating, sheared nebula flow, under the general condition $\Omega_L \ge \Omega_K$. The utility of the work is to allow for initial conditions in forced simulations of turbulence with particles that satisfy simultaneously the critical nebula parameters $\alpha$ and $\Omega_L$. Without actually knowing what the appropriate turbulent frequency $\Omega_L$ is, the relevant particle Stokes number St $\equiv t_s \Omega$ may be incorrectly defined -- here, $t_s$ is the aerodynamic stopping time of the particles. This appears to be the case in many prior simulations of particle behavior in nebula turbulence for which the particle properties are defined by an $\Omega_K$-based Stokes number \citep{Birnstiel_etal_2010, Estrada_etal_2016, Sengupta_etal_2019, Gole_etal_2020}, which we define as ${\rm St}_K=t_s \Omega_K$, and in which the possibility of  $\Omega_L > \Omega_K$ was not considered. In this paper we will introduce and utilize injection scale eddy Stokes number defined hereafter by ${\rm St}_L = t_s \Omega_L$.

\section{Generalizing the components of $\alpha$}\label{sec:ansatz}

\subsection{The current {\it ansatz} $\Omega_L = \Omega_K$}\label{sec:omegaL=omegaK}

\begin{figure*}
    \centering
    \includegraphics[width=\textwidth,]{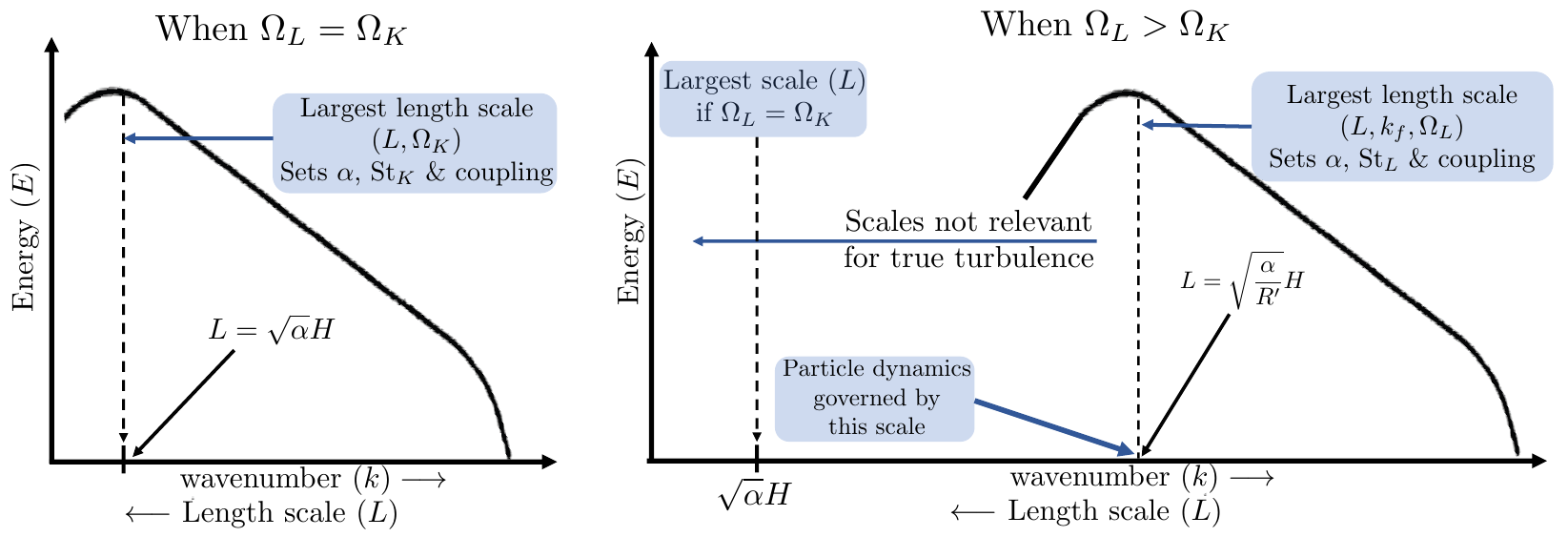}
    \caption{A cartoon showing  turbulent kinetic energy spectra $E(k)$ as a function of spatial wavenumber $k$ for two different cases: $\Omega_L=\Omega_K$ and $\Omega_L>\Omega_K$. Note that when $\Omega_L$ exceeds $\Omega_K$ the largest energy containing length scale of the turbulence shifts towards right. As this is the governing scale setting the turbulent dynamics, transport and diffusion properties get altered.} 
    \label{fig:cartoon}
\end{figure*}

We first revisit the original ansatz which is generally assumed to be true in literature in cases where the turbulence parameter is calculated from numerical simulations. We shall call this $\alphat$, the turbulence intensity when $\Omega_L=\Omega_K$.

\cite{Cuzzi_etal_2001} restated the separation of the $L$ and $V_L$ components of $\nu$ based on the ansatz that the largest scale eddies in the nebula would have eddy frequency $\Omega_L = \Omega_K$. Situations where $\Omega_L < \Omega_K$ generally corresponds to slow currents rather than fluctuating eddies, such as gently varying zonal flows.

These structures do not survive longer periods due to the Coriolis effect introduced by rotation \citep{Zahn_1989,Cuzzi_etal_2001}. 

When the ansatz $\Omega_L = \Omega_K$ is assumed true, $\alphat$ relates to the length and velocity scales in a specific way by starting with the usual formulation 
\begin{equation}\label{eqn:d_nu_t}
    (D,\nu) \sim LV_L \sim \alphat c H \sim \alphat H^2 \Omega_K,
\end{equation}
where $c$ is the local sound speed and we have used $H\equiv c/\Omega_K$, where $H$ is the local gas scale height. In this case, $c$ and $H$ are considered as the characteristic velocity and length scales of the system. Now, if the ansatz $\Omega_L = \Omega_K$ is valid, then Eq. (\ref{eqn:d_nu_t}) can be re-written as
\begin{equation}\label{eqn:d_nu_t_2}
    (D,\nu) \sim LV_L \sim L^2 \Omega_L \sim L^2 \Omega_K.
\end{equation}
Combining equations \ref{eqn:d_nu_t} and \ref{eqn:d_nu_t_2} gives $L^2 = \alphat H^2$ or $L=\sqrt{\alphat}H$. Substituting $L$ back in $LV_L = \alphat c H$ reveals
\begin{equation}
V_L=\sqrt{\alphat}c.
\label{VL-tildealpha}
\end{equation}
Often invoked in the form $\alphat=V_L^2/c^2$, this result is widely used to calculate particle relative velocities in growth models \citep[e.g.][]{Ormel_Cuzzi_2007, Birnstiel_etal_2010, Birnstiel_etal_2016,  Estrada_etal_2016, Sengupta_etal_2019}. The estimate $\alphat$
is also used as a measure of turbulent intensity in forced turbulence models, i.e., based on root mean square (rms) velocities \citep[][see section \ref{sec:SS} for more discussion]{Gole_etal_2020}.  


\subsection{The general case $\Omega_L > \Omega_K$}\label{sec:ansatz_not_true}

The above usual ansatz does not address cases where $\Omega_L > \Omega_K$, which may be the case in realistic turbulence.
In order to address cases where the traditional assumption $\Omega_L = \Omega_K$ is invalid \citep[as occurs in at least the highly relevant VSI regime;][]{Nelson_etal_2013,Stoll_Kley_2016}, the analysis of section \ref{sec:omegaL=omegaK} must be generalized. For the general situation we denote the turbulence (diffusion) parameter as $\alpha$ (rather than $\alphat$). Starting once again with Eq. (\ref{eqn:d_nu_t}), we write the frequency of the largest eddy as $\Omega_L\sim V_L/L$, where $V_L$ is the velocity of the largest (energy containing) eddy  and $L$ is the corresponding length scale. We define  $\Omega_L/\Omega_K \equiv R'$, where $R'$ is twice the traditional Rossby number \Ro \ \citep{Cushman-Roisin-Book}. 

With this, Eq. (\ref{eqn:d_nu_t_2}) now reads as
\begin{equation}
        (D,\nu) \sim LV_L \sim L^2 \Omega_L \sim R'L^2 \Omega_K 
\end{equation}
As before, setting this result equal to  Eq. (\ref{eqn:d_nu_t}) in terms of $\alpha$, we get
\begin{equation}\label{eqn:length}
    L=\sqrt{\frac{\alpha}{R'}}H. 
\end{equation}
Using Eq. (\ref{eqn:length}), the velocity $V_L$ for the largest eddy can be written as 
\begin{equation}
    \label{eqn:velocity}
    V_L=L\Omega_L=\sqrt{\frac{\alpha}{R'}}R'H\Omega_K=\sqrt{R'\alpha}c,
\end{equation}
which is equivalent to the expression
Eq. (\ref{VL-tildealpha}).
Thus,  while $L$ and $V_L$ change form, their product does not, and we recover
\begin{equation}
    \label{eqn:d_nu_t_final}
    (D,\nu) \sim LV_L = \alpha cH.
\end{equation}
From Eq. (\ref{eqn:velocity}), $\alpha$ can be related to $\alphat$ as
\begin{equation}\label{eqn:alpha_alphatilde}
    \alpha = \frac{V_L^2}{R'c^2} = \frac{\alphat}{R'}.
  \end{equation}
  In the limit of $\Omega_L=\Omega_K$,  $R'\rightarrow 1$ and the general case reduces to  $\alpha = \alphat$.
Also, substituting the above expression into Eq. (\ref{eqn:velocity}) demonstrates its equivalence
to Eq. (\ref{VL-tildealpha}).

It is noteworthy that a similarity can be drawn between the expression for the diffusive $\alpha$ found in Eq. (\ref{eqn:alpha_alphatilde})  and the one given by \citet[][see their section 3]{Youdin_Lithwick_2007}, i.e., $\sim \delta V_g^2 \tau_{{\rm eddy}}$, where $V_g$ is the rms of the gas fluctuation velocities and $\tau_{{\rm eddy}}$ is the eddy turnover time. Given that $\delta V_g^2$ is dominated by the largest energy containing scale, $\tau_{{\rm eddy}} \sim 1/R'$ if $\tau_K\sim 1/\Omega_K\sim 1$ is assumed\footnote{The interested reader is referred to \citet{Carballido_etal_2011} for a validation and discussion of the \citet{Youdin_Lithwick_2007} expression for particle diffusion.}.

\subsection{Why is $\Omega_L$ so pivotal in nebular turbulence?}
In the protoplanetary disk the gas scale height $H = c/\Omega_K$ is a natural length scale arising from hydrostatic balance. In a Keplerian disk it is also the largest scale on which organized motions like large vortices remain subsonic -- so if the turbulence reaches the scales of $H$ (through either that scale being unstable, or through inverse cascade, then $H$ will shape the turbulence).

The most significant length scale for turbulence is the scale associated with the largest eddy, the correlations of which govern major aspects of nebula physics from transport properties to particle dynamics, and it has been found that it can be orders of magnitude smaller than $H$ for many plausible turbulence mechanisms studied to date even when $R'=1$ (Figure \ref{fig:cartoon}, left). 

When the more general case is considered ($\Omega_L\ge\Omega_K$), the largest turbulent eddy length scale shifts towards smaller scales (Figure \ref{fig:cartoon}, right), and under such conditions it is this scale which dominates the particle transport and diffusion properties in the nebula. Scales longer than this may be important for zonal flows or coherent structures, such as long lived vortices, but are not relevant to fluctuating turbulent gas dynamics. 

\section{Simulations with Forced Turbulence.}\label{sec:cal_sims}

For the purpose of demonstration of the theoretical concept constructed in section \ref{sec:ansatz_not_true}, we executed a series of simulations where turbulence is forced externally in a shearing box with domain size $0.2H$ in radial and azimuthal direction and $0.4H$ in the vertical. All the simulations presented in this section are gas-only (Lagrangian particles are introduced in section \ref{sec:numericaltest}) and we solve for the equations for mass and momentum conservation under isothermal condition with vertical gas stratification in a shearing box setup:

\begin{equation}\label{eqn:gascontinuity}
    \frac{\partial \rho_g}{\partial t} + \nabla \cdot (\rho_g \bug_g) = 0;
\end{equation}
\begin{equation}\label{eqn:gasmomentum}
    \frac{\partial \bug_g}{\partial t} +\left(\bug_g \cdot \nabla\right)\bug_g 
    + 2\Omega_K {\bf{\hat z}}\times {\bf u}_g
    =-\frac{1}{\rho_g}\nabla P -\Omega_K^2 z \hat{{\bf z}} + 3\Omega_K^2 x{\bf {\hat x}},
\end{equation}
where, $\bug_g$ is the gas velocity, $P$ is the gas pressure, $\rho_g$ is the gas density and $\Omega_K$ is the local Keplerian frequency. The term $\Omega^2 {\bf z}$ in Eq. (\ref{eqn:gasmomentum}) represents the vertical gravity responsible for the gas stratification. The third term on the left hand side of equation \ref{eqn:gasmomentum} represents the coriolis force arising from the rotation and the last term of the right hand side of the same equation is the shear.

 The resolution used in all the simulations in this section is $(N_x\times N_y\times N_z) \equiv 256\times 256\times 512$. We have shear, rotation and vertical stratification included in the simulations, and use the {\sc Pencil Code}\footnote{\url{http://pencil-code.nordita.org/ }} \citep{Pencil_code} to solve the continuity and momentum equations. The forcing scheme is the same one used in \citet{Sengupta_Umurhan_2023}. The external forcing is controlled by a forcing parameter $f_0$ that sets the amount of energy injected in the system at a previously chosen wavenumber. For the gas-only simulations presented in this section, we have chosen two forcing wavenumbers  $k_f=3$ and $6$, and used $f_0 = [0.003, 0.005, 0.01, 0.03, 0.05, 0.1, 0.3]$ for each $k_f$, totalling $14$ simulations. The details of the forcing scheme are presented in appendix \ref{sec:forcing}. We have also used sixth-order hyperdiffusion and hyperviscosity in the simulations, allowing the fields to dissipate their energy near the smallest scales while preserving the power spectra at the larger scales \citep[for details see][]{Sengupta_Umurhan_2023}.  

We calculate $\alpha$ and $\alphat$ directly from our code results 

\begin{eqnarray}
\alpha  &\equiv& \frac{V_{{\rm rms}}^2}{R'c^2} \sim \frac{2}{3}\frac{\langle v_r^\prime v_\phi^\prime \rangle}{c_s^2}, \label{def:correct-alpha}\\
\alphat &\equiv& V_{\rm rms}^2/c^2 \sim V_L^2/c^2 \label{def:alphat_sims}
\end{eqnarray}

\noindent where $v_r^\prime$ and $v_\phi^\prime$ are the local velocity fluctuations relative to
the mean flow, and $V_{\rm rms}$ is a spatial and temporal average of the fluctuating velocity amplitudes. It is generally the case that $V_L \approx V_{\rm rms}$ (see Appendix \ref{sec:alpha_alphat} for details). 

\begin{figure*}
    \centering
    \includegraphics[width=\textwidth,]{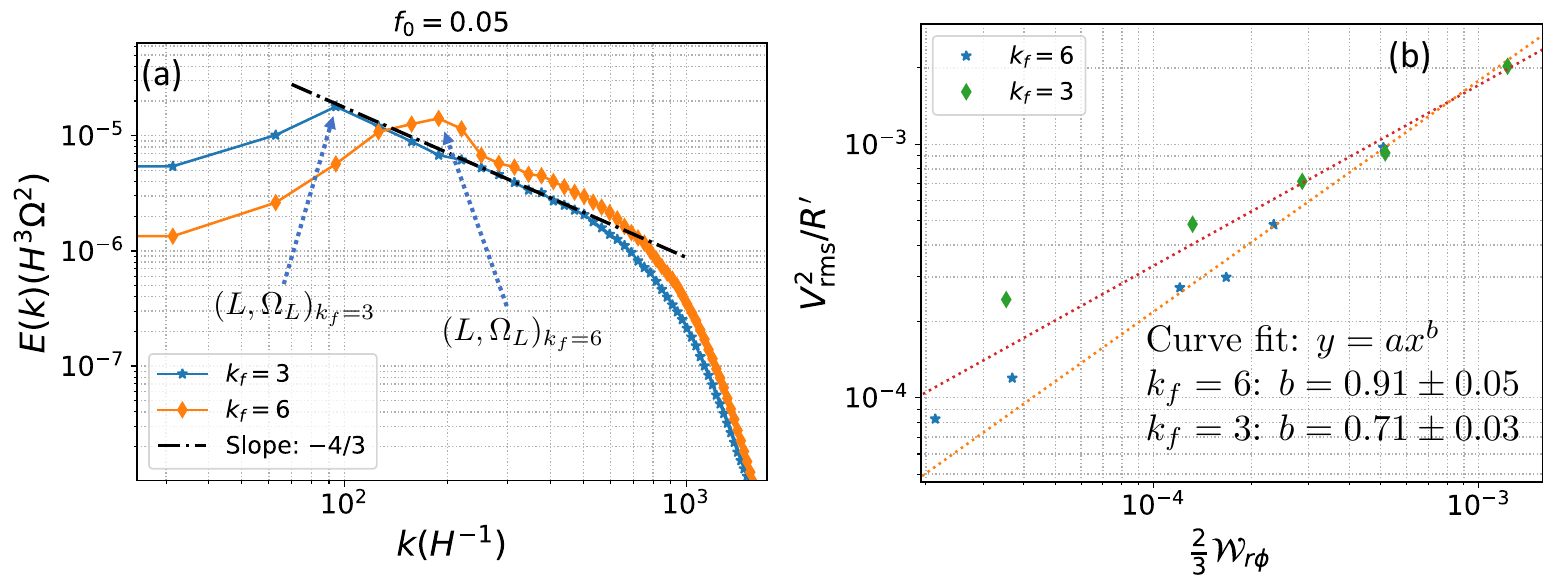}
    \caption{\textbf{a:} Kinetic energy spectra for two different wavenumbers: $k_f=3$ (blue star) and $6$ (orange diamond). The largest energy containing eddy coincides with the injection scales. Note the inertial range in both cases follow a $-4/3$ power law instead of $-5/3$  which  is due to the rotation of the system. \textbf{b:} The variation of $V_{{\rm rms}}^2/R'$ with the tangential Reynolds stress $\frac{2}{3}\tanstress$ and $R'$ calculated based on the KE spectrum and Eq. (\ref{eqn:omegaL}). The relationship shows a power-law with exponent $0.91\pm 0.05$ for $k_f=6$, close to our expectation of unity, and $0.71 \pm 0.03$ for $k_f=3$ (see section \ref{sec:SS} for details).}
    \label{fig:spectra}
\end{figure*}

\subsection{{The importance of turbulent kinetic energy spectra}}\label{sec:spectra}

Here we illustrate how relevant turbulence parameters are determined directly from the turbulent kinetic energy spectrum $E(k)$ as a function of spatial wavenumber $k$ (see figure \ref{fig:cartoon}). 
The first step to test the general case through our numerical simulations is to estimate $\Omega_L$. The velocity $\vk$ of an arbitrary eddy with wavenumber $k$ can be written as 
\begin{equation}\label{eqn:vk}
    \vk = \sqrt{2kE(k)},
\end{equation}
where $E(k)\,dk$ is the energy per unit mass contained in eddies with wavenumber ranging between $k$ and $k+\,dk$ where $E(k)$ has units of $H^3\Omega_K^2$. Then, the turnover time for eddies with wavenumber $k$ is
\begin{equation}\label{eqn:tau} 
\tau \sim \ell/\vk \sim \frac{1}{k\vk},   
\end{equation}
where $\ell \sim 1/k$ is the eddy length scale \citep{Tennekes_Lumley_1972, Davidson_2004}, and the corresponding eddy frequency is simply $\Omega_e \equiv 1/\tau$. Note that $\tau$ here is the same as the nonlinear timescale or the timescale associated with the scale-to-scale energy transfer.

Using Eqs. (\ref{eqn:vk}-\ref{eqn:tau}), the eddy turnover frequency can be expressed as 
\begin{equation}
    \label{eqn:turnoverfreq}
    \Omega_e \sim 1/\tau \sim k\vk=\sqrt{2k^3E(k)}.
\end{equation}
For the energetically dominant eddy with length scale $L$, Eq. (\ref{eqn:turnoverfreq}) reads as
\begin{equation}
    \label{eqn:omegaL}
    \Omega_L = \sqrt{2k_f^3E(k_f)},
\end{equation}
where $k_f$ is the energy injection scale or the forcing scale in our simulations.  Correspondingly, 
\begin{equation}\label{eqn:vL}
    V_L = \sqrt{2k_fE(k_f)},
\end{equation}
 with $L\sim 1/k_f$.  Figure \ref{fig:spectra}(a) illustrates these relationships for two energy spectra with $f_0=0.05$ with forcing at $k_f=3$ and $6$. Note the positions of the largest energy containing scales which determine $\Omega_L$ for the two cases.
{Appendix \ref{sec:mixing_length} shows how these spectral expressions also lead to $\alpha = \sqrt{\tilde\alpha}/k_f H =\tilde\alpha/R'$ (Eq. \ref{eqn:alpha_alphatilde}).}

\subsection{Connection between $\alpha$ and tangential Reynolds stresses}\label{sec:SS}
Following \citet{Shakura_Sunyaev_1973} the coefficient of turbulent dynamical viscosity $\mu$ is defined by the mixing length form 
\begin{equation}
     \mu \equiv \rho L V_L
\end{equation}
and using the correspondence between Reynolds stress $\mathtt{W}_{r\phi}$ and viscosity

\begin{equation}
    \mathtt{W}_{r\phi} = \mu  r\frac{\partial}{\partial r}\frac{v_\phi}{r} = -\frac{3}{2} \mu  \Omega_K,
    \label{shakura_sunyaev_stress_relationship}
  \end{equation}
in which $r$ is the cylindrical radius coordinate and where Keplerian flow is assumed 
for the azimuthal velocity $v_\phi$, i.e., $v_\phi/r = \Omega_K \sim r^{-3/2}$
\footnote{
In isotropic turbulence, the ensemble-averaged correlations of off-diagonal fluctuating velocity components are zero \citep{Davidson_2004}. However, in anisotropic systems like those involving Keplerian shear, off-diagonal velocity correlations like $\mathtt{W}_{r\phi}$ are not zero in general, and can lead to transport. This explains the original motivation behind the mixing-length model Eq (\ref{shakura_sunyaev_stress_relationship}).} 
Re-expressing the stress-viscosity relationship
Eq. (\ref{shakura_sunyaev_stress_relationship}) in terms of variables we use here for the box 
-- including the use of Eq. (\ref{eqn:d_nu_t_final}) and the assumption that the density $\rho$ is constant on the scales of interest -- we find
\begin{equation}
   \frac{2}{3} \tanstress = (\alpha c H)\Omega_K = \alpha c^2,
   {\label{SS_relationship}}
\end{equation}
where the above has been expressed in terms 
of the \textit{negative} of the specific tangential stress 
$\tanstress \equiv  -\mathtt{W}_{r\phi}/\rho$. 
Similarly to $V_{{\rm rms}}$, 
$\tanstress$ is calculated from the spatio-temporal average of the correlations between the radial and azimuthal velocity fluctuations (see Appendix \ref{sec:alpha_alphat}).   Eq. (\ref{SS_relationship})
simply states that turbulent kinematic viscosity  can be estimated from the specific tangential stresses via the simple form
$\alpha = 2\tanstress/3c^2$ (Eq.~\ref{def:correct-alpha}). However, the quantity $\tanstress$ (also commonly written as $<v_r' v_{\phi}' >$) is sensitive to the correlations of the velocity fluctuations in addition to their magnitudes, while $V_{{\rm rms}}$ is not, corroborating the analysis of \citet{Balbus_Papaloizou_1999}. Indeed in several cases, turbulent Reynolds stresses have been numerically calculated from actual velocity correlations rather than only their magnitudes \citep[e.g., as done for the VSI turbulence,][]{Stoll_Kley_2014, Nelson_etal_2013}. Such stresses can be directly associated with $\alpha$  \citep[also see][appendix B]{Youdin_Lithwick_2007}.

\par
This important distinction is  accounted for by our $R'$ scaling term as shown in Figure \ref{fig:spectra}. Using the $k_f=6$ simulation results in Figure \ref{fig:spectra}(a), Figure \ref{fig:spectra}(b) shows a fit of the form $y=ax^b$ between $\log (V_{{\rm rms}}^2/R')$ and $\log (2 \tanstress/3)$, where $b=0.91\pm 0.05$ - that is, $\tanstress \sim V_{{\rm rms}}^2/R'$. For $k_f=3$, however, we get $b=0.71\pm 0.03$ which is about $30\%$ less than the expected value. 

\begin{figure*}
    \centering
    \includegraphics[width=\textwidth,]{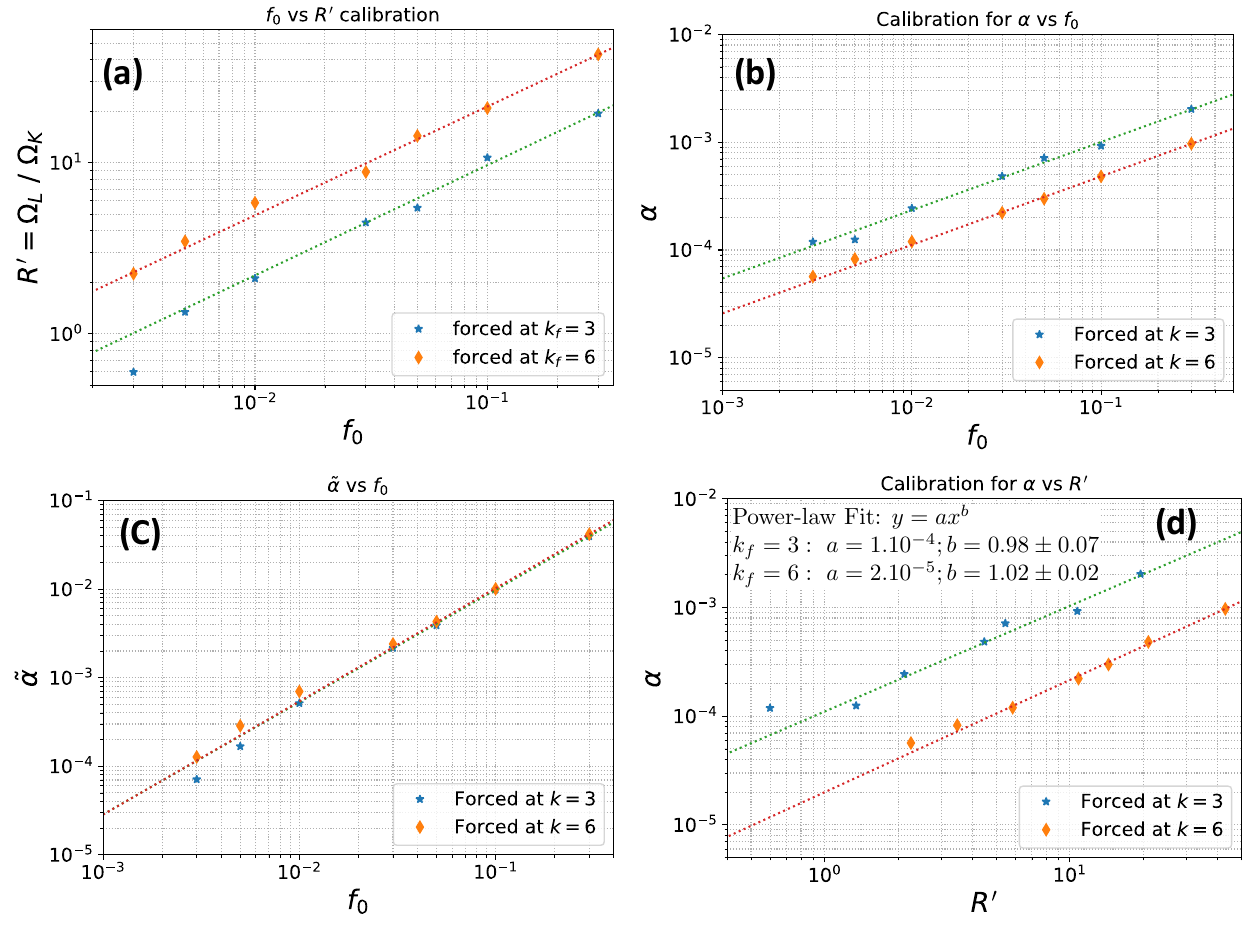}
    \caption{\textbf{a:} Calibration for $R'$ as a function of forcing energy injection, given in terms of the forcing parameter $f_0$ for two different injection wavenumbers: $k_f=3$ (blue star) and $k_f=6$ (orange diamond). The straight lines are the power-law fits for each data set. For fixed energy injection, $R'$ increases with $k_f$. \textbf{b:} Calibration for $\alpha$ (corrected by $R'$) vs $f_0$ for $k_f=3$ (blue star) and $k_f=6$ (orange diamond). \textbf{c:} Variation of $\alphat$ with $f_0$ for $k_f=3$ and $6$. Note that the values of $\alphat$ is almost the same for two different injection scales. With $\alphat\sim V{_{\rm rms}}^2/c^2$ over the full inertial range, $\alphat$ is expected to reflect the total energy injected into the system, irrespective of the injection scale. \textbf{d:} Variation of $\alpha$ with $R'$ for $k_f=3$ and $6$. A power-law fit for $y=ax^b$ gives $b=0.98\pm 0.07$ for $k_f=3$ and $1.02\pm 0.02$ for $k_f=6$ (see section \ref{sec:scaling}).}
    \label{fig:figure1}
\end{figure*}

\subsection{$\Omega_L$ in forced turbulence as a function of $f_0$}\label{sec:omegaL_FA}

In Figure \ref{fig:figure1}(a) we show the variation of $R'=\Omega_L/\Omega_K$ as a function of  $f_0$ for two different injection wavenumbers $k_f=3$ (blue star) and $6$ (orange diamond). For a particular forcing wavenumber $k_f$, $\Omega_L$ and hence, $R'$ increases with the forcing amplitude. Increasing the injection energy at the same wavenumber and associated length scale results in a faster rotation for the largest (energy containing) eddy and a corresponding increase in the turnover frequency $\Omega_L$. It is also evident from figure \ref{fig:figure1}(a) that for a fixed value of $f_0$, $R'$ increases when energy is injected at a higher wavenumber and smaller scale, as for fixed energy injection, smaller eddies have a smaller turnover time.

\subsection{$\alpha$ in forced turbulence as a function of $f_0$.}\label{sec:alpha_f0}

In Figure \ref{fig:figure1}(b), a calibration of $\alpha = \alphat/R'$ as a function of $f_0$ is shown for forcing at two different wavenumbers: $k_f=3$ (blue star) and $6$ (orange diamond). It is evident from the figure that for a fixed $k_f$, $\alpha$ increases with the increasing energy injection. Diffusion and viscosity are primarily governed by the largest energy containing scales, so increasing energy at those scales will end up with an increased diffusion and hence, a higher $\alpha$. However, for a fixed $f_0$, $\alpha$ decreases when energy is injected at a higher wavenumber (see $k_f=3$ vs $k_f=6$ plots in Figure \ref{fig:figure1}(b) as the length scale for the largest energy containing eddy decreases when a higher wavenumber is chosen.

Interestingly, as $\alphat$ is not a function of $R'$ and depends on the rms of the fluctuating velocities throughout the inertial range, which measures the total energy injected into the system, variation in $\alphat$ for a fixed $f_0$ is minimal with the injection wavenumber $k_f$ as shown in Figure \ref{fig:figure1}(c). This result is also consistent with our findings in Eqs. (\ref{eqn:velocity}) and (\ref{eqn:alpha_alphatilde}) that $V_L^2$ remains reasonably constant with the forcing wavenumber and $R'$ implicitly brings the length scale information in the realization of $\alpha$.

\subsection{Scaling between $\alpha$ and $R'$}\label{sec:scaling}

The present model, developed on the basis of the general case $\Omega_L > \Omega_K$, can be tested by analysing the scaling between $\alpha$ and $R'$, using the  parameter $f_o$. Consider that the total energy injected varies with $f_0$ with some power-law index $p$ as $E(k)\propto f_0^p$. With $R'=\Omega_L/\Omega_K$ and $\Omega_L\propto E(k)^{1/2}$ as in Eq. (\ref{eqn:omegaL}), we can write $R'\propto E(k)^{1/2}\propto f_0^{p/2}$. Similarly for $\alpha$, following Eq. (\ref{eqn:alpha_sengupta}), $\alpha = V_L^2/R'$ and $V_L\propto E^{1/2}$, as in Eq. (\ref{eqn:vk}); thus we can write $\alpha\propto E^{1/2}\propto f_0^{p/2}$. So, we expect that for a particular value of $k_f$, $\alpha/R'$ is constant and the slope of the plot for $\log \alpha$ vs $\log R'$ will be unity.

In figure \ref{fig:figure1}d, we show the variation of $\log\alpha$ vs $\log R'$ for $k_f=3$ (blue star) and $6$ (orange diamond), with the calibrations shown by dotted lines. When a power-law fitting in the form $y=ax^b$ is done for the respective data sets, we get $b=0.98\pm 0.07$ for $k_f=3$ and $b=1.02\pm 0.02$ for $k_f=6$, 

\begin{figure*}
    \centering
    \includegraphics[width=\textwidth]{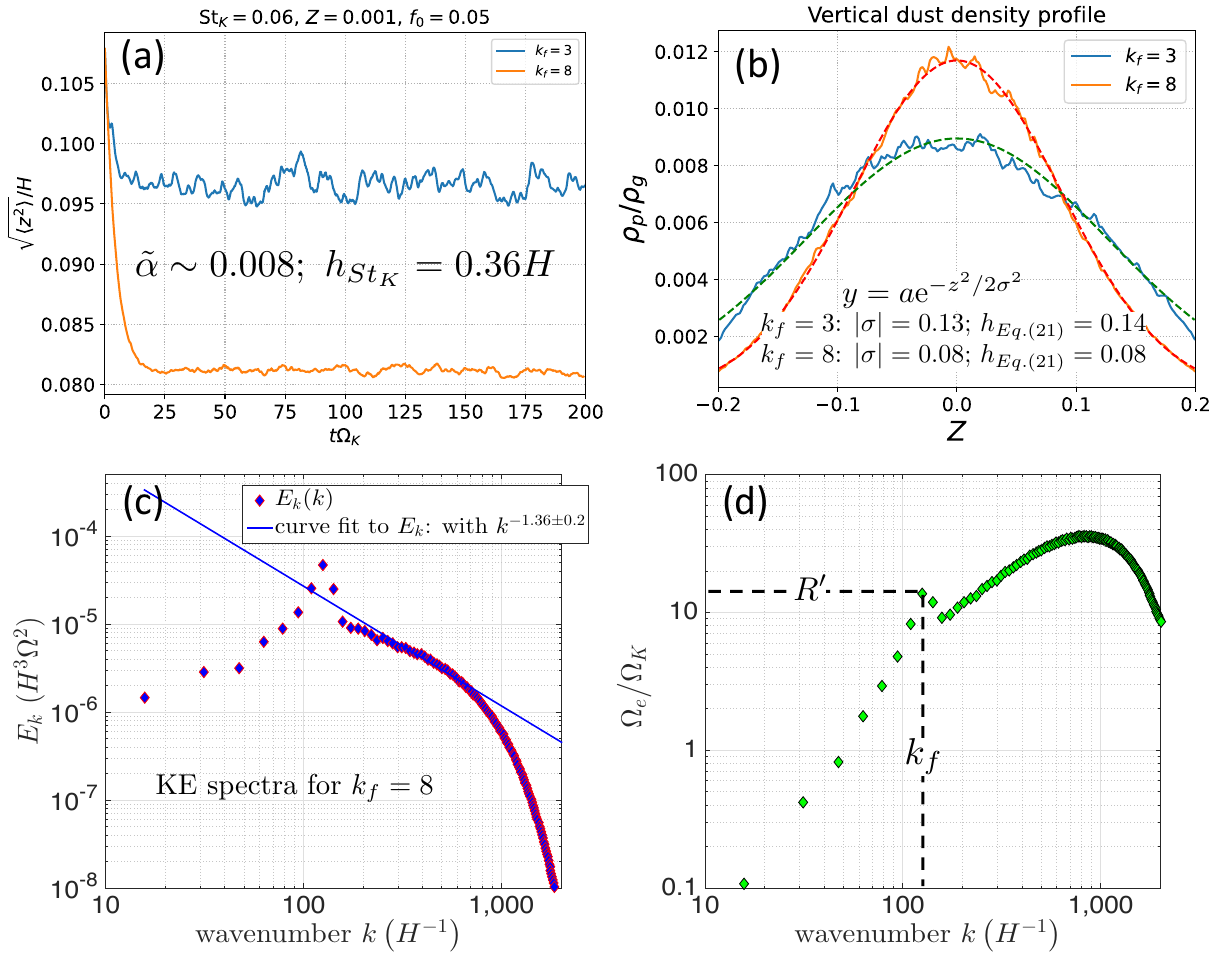}
    \caption{ \textbf{a:} Scale heights $h$ for particles with St$_K=0.06$ when forcing at $k_f=3$ (blue dashed curve) and $8$ (orange solid curve). Here $f_0=0.05$ producing $\alphat\sim 8\times 10^{-3}$, and $R'\sim 5.5$ and $\sim 12$ for $k_f=3$ and $8$ respectively. The St$_L$ for the two wavenumbers are $0.3$ (for $k_f=3$) and $0.7$ (for $k_f=8$). The simulations show $\zrms \sim 0.097H$ and $0.082H$ for $k_f=3$ and $8$ respectively. The scale height calculated using St$_K$ and $\alphat$ gives $h=0.36$. \textbf{b:} The vertical particle density distribution for two injection scales and the respective Gaussian fits in the form $y=a~{\rm exp}(-x^2/2\sigma^2)$ on the top. The fitting gives the values of the parameter $|\sigma|=0.13\pm 0.0007$ for $k_f=3$ and $0.08\pm 0.0001$ for $k_f=8$. The particle distribution is Gaussian with the midplane particle-to-gas mass ratio reaching $\sim 0.009$ for $k_f=3$ and $\sim 0.012$ for $k_f=8$   implying that the effect of particle mass loading is negligible. \textbf{c:} Kinetic energy spectra for the particle-gas simulation forced at $k_f=8$ ($k\sim 120$ when normalized by domain size). A curve fit for the slope of the inertial range gives $-1.36\pm 0.2$, shallower than $-5/3$ even in the presence of inertial particles.  \textbf{d:} A plot for eddy turnover time $\Omega_e/\Omega_K$ for all the wavenumbers. The location for the largest energy containing wavenumber $k_f$ and the corresponding value of $R'$ are noted by the dashed straight lines.}
    \label{fig:scaleheight}
\end{figure*}

\subsection{Physical processes leading to different $R'$}
The model developed in this article is relevant to several kinds of nebula turbulence, such as that driven by vertical shear instability \citep[VSI;][]{Nelson_etal_2013, Flock_etal_2020} or Convective Overstability  \citep[COV;][]{Lyra_2014}. For example, \citet{Stoll_Kley_2014} found that their VSI simulations had $\Omega_L$ significantly higher than $\Omega_K$ (eddy turnover time $\tau_{{\rm eddy}}<1)$. More recently, the energy spectra presented for global VSI simulations in \citet{Flock_etal_2020}  show that energy injection happens around $k\sim 100$ and the stratification induced inverse cascade sets the largest energy containing scale around $k\sim 20$, suggesting that the value of $R'$ may significantly exceed unity. More recently, \citet{Sengupta_Umurhan_2023} found \Ro$>1$  in particle laden disk midplane when turbulence is driven by Kelvin-Helmholtz and symmetric instability \citep{Stamper_Taylor_2017}.  If the disk turbulence is primarily triggered and sustained by such hydrodynamic processes, the method of approximating $\alpha$ as $\sim \delta v^2/c^2 (= \alphat)$ will be an over-estimation, whereas values for $\alpha$ from the analyses of edge-on disks will probably be more consistent with the actual level of disk turbulence (see section \ref{sec:hd_observation}).

\section{Numerical Validation}\label{sec:numericaltest}

There are several situations in which the value of $R'$ becomes relevant -- particle scale height being one of them. In order to investigate the validity of $\Omega_L > \Omega_K$ and the effect of $R'$, we have performed numerical tests where inertial $($St$\neq 0)$ particles have been introduced in the simulation box with rotation and vertical stratification, with the turbulence being generated and maintained by external forcing. The particles are then allowed to settle under the influence of vertical gravity until a balance between the settling and turbulent diffusion is reached. We then investigated the scale heights obtained in the simulations, the main objectives being testing whether inertial particles react to the nebula turbulence following $(\alpha, \rm{St}_L)$ or $({\alphat}, \rm{St}_K)$.

In a protoplanetary disk, a steady state between particle settling towards the disk midplane and turbulent diffusion in the vertical direction sets the particle scale height \citep[also see Appendix \ref{sec:hpdetails}]{Cuzzi_etal_1993, Dubrulle_etal_1995, Youdin_Lithwick_2007}:
\begin{equation}\label{eqn:scaleheight}
    h = H \sqrt{\frac{\alpha}{\rm{St}_K(1+\rm{St}_L^2)}}.
\end{equation}

In order to test the validity of the theory and insights developed in sections \ref{sec:ansatz} and \ref{sec:cal_sims}, we now introduce Lagrangian particles in our simulations and compare their scale heights $h$  with Eq. (\ref{eqn:scaleheight}). We use a monodisperse particle distribution with St$_K = 0.06$ and an initial vertically averaged particle-to-gas mass ratio $Z=0.1\%$. Thus while the particle back-reaction is formally in play, the particle local density never gets large enough for it to be significant. The particles are coupled to the gas through aerodynamic drag 
by their stopping time $t_s$ (section \ref{sec:introduction}) - the time required for a particle to lose all its momentum to the differentially moving gas. 
Here, the Lagrangian equations of motion for a particle's vector position ({\bf{$x_p$}}) and velocity ({\bf{$u_p$}}) can be written as


\begin{equation}\label{eqn:dustcontinuity}
\frac{d{\bf{x_p}}}{dt}=\bf{u_p},
\end{equation}
\begin{equation}\label{eqn:dustmomentum}
\frac{d{\bf{u_p}}}{dt}
+ 2\Omega_K {\bf{\hat z}}\times {\bf u_p}
=-\Omega_K^2 z {\bf {\hat z}} +3\Omega_K^2 x{\bf {\hat x}} -\frac{\bup -\bug_g({\bf{} x_p})}{t_s}.
\end{equation}
The second term on the right hand side in Eq. (\ref{eqn:dustmomentum}) represents momentum exchange via aerodynamic drag that depends upon the gas velocity evaluated at the particle's position.  In the gas momentum equation Eq. (\ref{eqn:gasmomentum}), we must introduce a term representing the equal and opposite momentum transfer. This modifies the Eq. (\ref{eqn:gasmomentum}) to:
\begin{eqnarray}
  & &   \frac{\partial \bug_g}{\partial t} +\left(\bug_g \cdot \nabla\right)\bug_g 
    + 2\Omega_K {\bf{\hat z}}\times {\bf u}_g
    = \nonumber \\
  & & \ \ \   -\frac{1}{\rho_g}\nabla P -\Omega^2 z \hat{{\bf z}}   +3\Omega_K^2 x{\bf {\hat x}}
    -\frac{\rho_p}{\rho_g}\frac{\bug_g - \bup}{t_s},
\end{eqnarray}
where $\rho_p$ is the local volume mass density of the particle phase.

In Figure \ref{fig:scaleheight}, we present the results of our numerical experiment where turbulence is forced with $f_0=0.05$ at $k_f=3$ (blue dashed curve) and $8$ (orange solid curve) with $k\sim 45$ and $120$ respectively when normalized by domain size, producing $\alphat\sim 8\times 10^{-3}$ in a simulation box, including vertical gravity, rotation and density stratification. The vertically integrated solid to gas mass ratio in the simulation is kept low ($Z=0.1\%$) in order to minimize the effects of particle mass loading. The scale height for the particles when calculated using $\alphat$ and St$_K$ is $h=0.34H$, significantly larger than what the simulations show. In Fig. \ref{fig:scaleheight}(a) the obtained $\zrms$ is shown which is different for the two runs in spite of all the parameters being the same and differ only by the forcing wavenumber, clearly indicating that $R'$ plays an important role in setting $\alpha$ and hence, the vertical distributions of the particles. The values of $R'$ for $k_f=3$ and $8$  are approximately $5$ and $12$, which produces St$_L \sim 0.3$ and $0.7$ respectively. When $h$ is calculated using Eq. (\ref{eqn:scaleheight}) with these values, we get $h\sim 0.14$ for $k_f=3$ and $\sim 0.08$ for $k_f=6$. In Fig. \ref{fig:scaleheight}(b) the particle density for the two simulations are shown with a Gaussian fit in the form $y=a{\rm e}^{-z^2/2\sigma^2}$ on the top. The fitted values for $|\sigma |$ (the particle scale height of the distribution) for the two wavenumbers come out to be $0.13$ and $0.08$, a close match to the calculated scale heights from Eqn. (\ref{eqn:scaleheight}).
So, it is evident from our numerical experiment that the vertical diffusion of particles under nebula turbulence should be effectively parametrized by $\alpha$, not $\alphat$, and without the scaling by $R'$ as in Eqs. (\ref{eqn:alpha_alphatilde}) and (\ref{eqn:alpha_sengupta}), the reported $\alphat$ will be an overestimation of the true diffusivity of the particles in the domain. 

We calculate $R'$  using equation Eq. (\ref{eqn:omegaL}) directly from the KE spectra produced from the simulation data. In Fig. \ref{fig:scaleheight}(c) the kinetic energy spectra for the particle-gas simulation with $k_f=8$ is shown. The associated eddy turnover times $\Omega_e$ for all wavenumbers are shown in Fig. \ref{fig:scaleheight}(d). The vertical and horizontal dashed black lines in the figure indicates the forcing wavenumber (length scale $L$) and the associated $\Omega_L = R'\Omega_K$.

Note that the estimated particle scale heights $h$ using $\alpha$ and St$_L$ match nicely with the values for $|\sigma|$ obtained from the Gaussian vertical particle distribution. However, when compared with $\zrms$, the estimate agrees well with $k_f=8$ but differ by $\sim 35\%$ for $k_f=3$. This difference is due to the effect of a relatively small vertical domain used in the simulation. $h$ gets smaller either when the level of turbulence is low or the energy is injected at a higher wavenumber (smaller scale) so that the velocity correlations at the largest energy containing scales are small. This is the reason why $\zrms$ agrees well with the actual scale height for $k_f=8$. We also expect that $\zrms$ and $|\sigma|$ will be close for $k_f=3$ if a smaller value for $f_0$ (lower level of turbulence) where particles layer thickness is small, or a larger vertical domain is used. So, we can infer that $\zrms$ is a good representation of particle scale height only when the particles are settled sufficiently away from the vertically periodic boundary of the simulation domain.

\section{Discussion}

\subsection{Determination of $\alpha$ from observations.}\label{sec:hd_observation}

There are several situations in which the value of $R'$ is relevant -- estimating the level of nebula turbulence from disk observations is one of them. Observations of protoplanetary disks reveal that the systems are turbulent to varying degrees and the observed levels differ by the mode of observation as well.  

One way of assessing $\alpha$ for disks through observations is by analysing the turbulent line broadening in order to obtain $\delta v/c$ and estimating $\alpha$ as $\sim \delta v^2/c^2$ using velocity information only \citep{flaherty_etal_2017, Flaherty_etal_2018, Flaherty_etal_2020, Teague_etal_2018}. Generally this method gives an upper bound of $\alpha$ -- for example, \citet{flaherty_etal_2017} found $\delta v/c\sim 0.01$ for HD 163296 and $\sim 0.3$ at $\sim 1-2$ scale height above the midplane for DM Tau (Also see \citet{Lesur_etal_2022} for a discussion). Similar analysis by \citet{Teague_etal_2018} on TW-Hya using the $J=7-6,~5-4$ and $3-2$ transitions of CS imaged at $\sim 0.5''$ spatial resolution estimated $\delta v/c\lesssim 0.1$. As these observations improve, it will be critical to remember that they are fundamentally $V_{{\rm rms}}^2$-based and thus measure $\alphat$, not $\alpha$. 

On the other hand, the observed vertical thicknesses of edge-on disks with ALMA and the VLA have been used to infer values of $\alpha/$St using Eq.(\ref{eqn:scaleheight}). 
Then, values of $\alpha$ are inferred using assumptions of St based on these observations and nebula growth models \citep{Doi_Kataoka_2021, Doi_Kataoka_2023, Villenave_etal_2022, Villenave_etal_2023}. 
In this way, \citet{Villenave_etal_2022} estimated $\alpha\sim 10^{-5}$ for Oph 163131.
For HD 163296, which is an older Class II disk, estimates for $\alpha$ vary based on which ring system is being analyzed: for the inner more vertically extended ring the turbulence is vigorous with $\alpha = 6\times 10^{-2}$ \citep{Doi_Kataoka_2023}, while the corresponding values for the outer flatter ring are systematically lower but with little agreement between published values with $\alpha \le 6\times10^{-3}$ \citep{Doi_Kataoka_2023}, or
$\alpha \approx 4.5\times 10^{-3}$
\citep{Pizzati_etal_2023}.

We underscore that particle layer thicknesses are sensitive to $\alpha$, not $\alphat$ (section \ref{sec:numericaltest} and figure \ref{fig:scaleheight}). It is interesting to note that, while uncertainties remain significant (especially in St),  the estimated $\alpha$ from the analyses of particle scale heights seem to be consistently smaller compared to the ones obtained from the turbulent line broadening, and this is just what would be expected from the difference between values of  particle height derived from $\alphat$ and $\alpha$ if $R' > 1$.

It is possible to combine the  independent and distinct observational constraints discussed above (on $h/H$ from particle layer thicknesses and $V_{rms}$ from spectral line observations) with other observed constraints on particle size and gas disk surface mass density \citep[eg.,][]{Carrascogonzalez_etal_2019,Macias_etal_2021, Doi_Kataoka_2023} to actually constrain the critical value of $R'$ and thus the nature of the underlying processes causing disk turbulence in the first place. We sketch the logic by which this may be done in Appendix \ref{sec:revised_alphaSt_from_obs}.

\subsection{$R'$ and Sc: effects on global particle redistribution}\label{sec:schmidt}
 Here we comment on how $R'$ may affect particle radial distributions. As noted in section \ref{sec:introduction}, the ratio $\nu/D \equiv$ Sc (the Schmidt number) is usually taken as unity; indeed there has been some confusion in the literature as to the definition of Sc itself \citep[see \citet{Jacquet_etal_2012} and ][their Appendix B]{Estrada_Cuzzi_2016}.  As shown in section \ref{sec:SS}, this quantity is the same as the ratio $<v_r'v_{\phi}'>/V^2_{\rm rms} = 1/R'$. To our knowledge \citet{Hughes_Armitage_2010, Hughes_Armitage_2012} represent the most complete (or only) study of the role of variable Sc in global redistribution of particles undergoing growth, drift, and diffusion. They only varied Sc from 0.5-2.0, and over this range found the effects minor. Unsurprisingly, (what we find as) the more realistic case Sc$<$1 favors diffusion over drift or advection, leading to a slightly greater retention of small particles in outer disks after longer times. To the degree that future studies of different kinds of nebula turbulence might be found to favor larger values of $R'$, the implications for particle radial transport might need to be reassessed. This would include studies where the radial structure of a particle band is used to infer the properties of the local particle size and/or $\nu$ \citep{Dullemond_etal_2018}.
 
\vspace{0.1 in}
\subsection{Effects on Collision Velocities and Particle Growth}\label{sec:collision_growth} 

Following \citet{Voelk_etal_1980}, \citet{Ormel_Cuzzi_2007} derived closed-form solutions for the collision velocities for particles in the turbulent nebula for three different regimes: (1) the tightly coupled particles when $t_s<t_{\eta}$, where $t_{\eta}$ is the stopping time at the Kolmogorov scale (their equation 27); (2) the inertial range regime where $t_{\eta}<t_s<t_L$ (their equation 28); and (3) the heavy particle limit when $t_s>t_L$ (their equation 29). The response of a particle to turbulence is determined by ${\rm St}_L$, which is a factor of $R'$ larger than the ${\rm St}_K$ assumed by previous studies along these lines \citep[e.g.,][]{Zsom_etal_2010, Birnstiel_etal_2010,  Estrada_etal_2016, Sengupta_etal_2019, Sengupta_etal_2022}. This has different implications in the three different particle size ranges, as noted below. Regarding the equations below, recall that \citet{Ormel_Cuzzi_2007} {\it defined} St with the {\it meaning of} St$_L$, while at the time implicitly associating it with St$_{\rm K}$ as is done by all dust growth models till date.

The velocities for the tightly coupled small particles in the sub-Kolmogorov regime are mostly dominated by Brownian motion \citep[][see their figure 2]{Sengupta_etal_2019} and are unlikely to change significantly. However, with the $R'$ effect included, the collision velocities $\Delta V_{12}$ between particles with large-eddy Stokes numbers St$_{1L}$ and St$_{2L}$ \citep[Equation 27][]{Ormel_Cuzzi_2007} for the tight coupling limit becomes
\begin{equation}\label{eqn:regime1}
    \Delta V_{12}^2 = R'^2 V_g^2 ({\rm St}_{1{\rm K}} - {\rm St}_{2{\rm K}})^2
\end{equation}

Similarly, in the most important range (2) above, Eq. (28) of \citet{Ormel_Cuzzi_2007} shows the collision velocity, which can be written with the new scaling included as
\begin{equation}\label{eqn:regime2}
    \Delta V_{12}^2 = R' V_g^2 \Lambda {\rm St}_{1{\rm K}}
\end{equation}
where the term $\Lambda$ is only weakly dependent on St$_L$ for St$_L \ll 1$ \citep{Ormel_Cuzzi_2007}. 
Thus, Eq. (\ref{eqn:regime2}), along with Eq. (\ref{eqn:velocity}), shows that the collision velocity is increased by a factor of $R'^{1/2}$ for a given $\alpha$, and thus particles reach the bouncing or fragmentation barriers at physical sizes  which are smaller by a related factor. 
  
 In the heavy particle limit, a higher ${\rm St}_L$ with $R'>1$ reduces collision velocities, allowing for bigger particles.
 The modified expression from \citet[][Eq. 29]{Ormel_Cuzzi_2007} would read as:
 \begin{equation}
      \Delta V_{12}^2 = V_g^2 \left(\frac{1}{1 + R' {\rm St}_{1{\rm K}}} + \frac{1}{1 + R' {\rm St}_{2{\rm K}}}\right)
 \end{equation}
 
\subsection{Planetesimal Formation}\label{sec:planetesimal}

Our results have implications for the process of planetesimal formation, an outstanding problem in  planet formation theory. 

For example, we have seen that turbulence with $R'>1$ can significantly reduce the particle scale height (figure \ref{fig:scaleheight}) and increase $\epsilon = \rho_p/\rho_g$ at the midplane, for particles of some given ${\rm St}_K$ or nominal physical size, relative to estimates based on the rms velocity-based turbulence parameter $\alphat$. This is primarily because the (more appropriate) value of $\alpha$ will decrease, and because $\rm{St}_L$ for the particles will increase (however, equation \ref{eqn:scaleheight} shows the St$_L$ effect to be relatively minor when ${\rm St}_L \ll 1$).   Planetesimal formation by either Streaming Instability  \citep[SI;][]{Youdin_Goodman_2005, Johansen_etal_2007} or Turbulent Concentration \citep[TC;][]{Cuzzi_etal_2010,Hartlep_Cuzzi_2020} is aided when the near-midplane $\epsilon$ is larger; even though this effect may only increase $\epsilon$ by a factor of order unity, this can be important. 

 TC has a more subtle dependence on ${\rm St}_L$ than SI, but comparison of planetesimal formation models with observations of ({\it e.g.)} observed chondrule or aggregate sizes will be affected. \citet{Hartlep_Cuzzi_2020} find reasonable sizes and abundances for their resulting planetesimals if the initial pebbles are centimeter and decimeter-sized, implicitly assuming $R'=1$. However, if $R'\sim 10$ for example, the same outcome would be achieved with mm-cm sized particles.

\section{Conclusions}

 The main objective of the work presented here is to investigate how the particle behavior in a turbulent fluid under nebular conditions is altered with varying large eddy frequency $\Omega_L > \Omega_K$. Below, we present a summary of our findings:

 \begin{itemize}
     \item When $\Omega_L>\Omega_K$, the characteristic length and velocity scales for  nebula turbulence are $L=\sqrt{\alpha/R'}H$ and $V_L=\sqrt{\alpha R'}c$, where $H$ and $c$ are the gas scale height and local sound speed respectively, $R'=\Omega_L/\Omega_K$, and $\alpha$ is the $<v_r' v_{\phi}' >$-like true gas kinematic viscosity parameter. 
     \item Under our more general scenario, the turbulent $\alpha$ of the nebula scales as $\alpha = \alphat/R'$, where $\alphat \sim V_{{\rm rms}}^2/c^2$ is the turbulence parameter when $\Omega_L=\Omega_K$. The parameter $\alphat$ is commonly used as a convenient approximation for the correct value $\alpha \propto \langle v_r^\prime v_\phi^\prime \rangle$, up to an undetermined correction factor, supposedly of order unity. Here we quantify this correction factor to be  $R'$, which can be significantly larger than unity. In this context, it is important to note that the calculation of $\alpha$ from the correlations of velocity fluctuations always incorporates the $R'$ dependence and is itself a proper measure of diffusivity or viscosity. 

     \item The effective particle Stokes number is best defined by the large eddy frequency $\Omega_L$ rather than the orbital frequency $\Omega_K$. When $\Omega_L>\Omega_K$,  ${\rm St}_L=R'{\rm St}_K$. St$_L$ is the parameter that governs the particle transport properties 
     and should be used for turbulent concentration simulations when comparing with the cascade model predictions from {\it eg.,} \citet{Hartlep_Cuzzi_2020}.
     \item When $R'>1$ ($\Omega_L>\Omega_K$), the particle layer scale heights $h$ in our 3D simulations show excellent match with the predictions obtained from ($\alpha, {\rm St}_L$) instead of from $(\alphat, {\rm St}_K)$.
     \item The self-consistent generation of turbulence by hydrodynamic processes, such as vertical shear instability \citep{Nelson_etal_2013} consistently shows $R'>1$ \citep{Stoll_Kley_2016, Flock_etal_2020}. Our findings imply that particle scale heights inferred from  turbulent broadening $(\alphat \sim \delta v^2/c^2)$ will be overestimates, while those obtained from direct observations of particle scale heights in edge-on disks or actual calculated Reynolds stresses are diffusion-based and probably more appropriate. 

     \item The collision velocity of particles, under the condition $\Omega_L>\Omega_K$, is scaled by different positive powers of $R'$ suggesting that for $R'>1$ turbulence, the maximum particle sizes should be smaller compared to $R'=1$ turbulence, all other things being equal. 
     \item The reduction of particle layer scale height and the enhancement of midplane $\epsilon =\rho_p/\rho_g$ for high $R'$ turbulence in the nebula should favor planetesimal formation by both Streaming Instability \citep{Youdin_Goodman_2005, Johansen_etal_2007} and Turbulent Concentration \citep{Hartlep_etal_2017, Hartlep_Cuzzi_2020}.
     \item Calibration of a numerical code for the purpose of particle-gas simulations under external forcing should be done separately for different particle sizes, as the mass loading changes the values for $\alpha$ and $R'$ for same injection energy. Furthermore, for external forcing in the presence of vertical stratification, choosing a forcing scale away from the domain size is a safer choice and is advisable (preferably $k_f=4$ or higher).
 \end{itemize}

In conclusion, we would like to emphasize the importance of turbulent kinetic energy spectra in the context of numerical studies of $\alpha$, and in particular, directly determining the value of $\Omega_L$ based on these spectra.

\section*{Acknowledgements:} We are grateful to Karim Shariff and Uma Gorti for multiple valuable conversations. We also thank Neal Turner for a careful review of the manuscript and important suggestions that significantly improved the manuscript. We thank our reviewer for comments which improved the quality of the paper. DS acknowledges support from NASA Prosdoctoral Program (NPP) Fellowship, NASA Astrobiology Institute, NASA Theoretical and Computational Astrophysical Networks (TCAN) via grant 80NSSC21K0497 and NSF via grant AST-2007422. All the simulations presented in this paper are performed on the NASA Advanced Supercomputing  (NAS) facility with generous computational resources provided through NPP, ISFM and TCAN allocations.

\newpage
\appendix

\section{Physics of the $\Omega_L =\Omega_K$ ansatz}\label{sec:Shakuraetal1978}
While the $\alpha$-model was introduced in \citet{Shakura_Sunyaev_1973}, they somewhat casually suggested that $V_L=\alpha c$, and this simple suggestion appears in a number of early papers. However, the underlying physics of the more commonly used form $V_L=\sqrt{\alpha} c$, which section \ref{sec:omegaL=omegaK} shows rests on an assumption that $\Omega_L =\Omega_K$, can be found in Appendix A of \citet{Shakura_etal_1978}. In particular their equations A1 and A2 give the energy dissipation rates $\dot{E}$ of the large eddies, and of the global nebula under Keplerian shear, respectively (in our notation) as
\begin{equation}\label{eqn:S78_eddies}
\dot{E}_{{\rm ed}}= V_L^3/L = V_L^2\Omega_L
\end{equation}
and 
\begin{equation}\label{eqn:S78_keplerian}
\dot{E}_{{\rm Kep}}= \nu\left(R \frac{\partial \Omega_K}{\partial R} \right)^2 = \frac{9}{4}\nu \Omega_K^2 \sim \frac{9}{4} L V_L \Omega_K^2 .
\end{equation}
Setting these energy dissipation rates equal leads directly to $\Omega_L = \Omega_K$, and represents the closure relation needed to separate $L$ and $V_L$. The ansatz is thus underlain by the physically plausible assumption that the turbulent kinetic energy budget of the large eddies (eqn A1) is supplied by the local release of gravitational energy by inwardly drifting mass, which is in turn caused by viscous outward transport of angular momentum \citep{LyndenBell_Pringle_1974}. The gravitational energy released is locally transformed to mechanical turbulent motions by viscosity acting across the local Keplerian gradient (eqn. A2). In all this of course, `viscosity' is generalized to a turbulent viscosity. This turbulent kinetic energy budget is transformed to heat on the smallest scales of turbulence and radiated away locally.  
On this basis, \citet{Shakura_etal_1978} then used a logic parallel to section \ref{sec:omegaL=omegaK}, which they expressed as $V_L/c \equiv {\cal M} \sim L/H \equiv \delta$, where ${\cal M}$ is a Mach number. In their section 3 they had already found that $\alpha \sim {\cal M} \delta \sim {\cal M}^2 \sim \delta^2$, leading directly, as in section \ref{sec:omegaL=omegaK}, to $V_L=c \sqrt{\alpha}$ and $L=H \sqrt{\alpha}$. 

Extending this ``energetics" logic to physical conditions  where other nebula structures are driven by non-keplerian shear, and have different local energy dissipation rates,  it's not hard to see how the large, energy-containing eddies can have $\Omega_L \ne \Omega_K$. For example, \citet{Cuzzi_etal_1993} studied midplane turbulence driven by a settled particle layer, which imposes a vertical velocity shear considerably larger than the radial Keplerian shear,  recently demonstrated explicitly in \citet{Sengupta_Umurhan_2023}. \citet{Cuzzi_etal_1993}  argued that the largest eddies would have large $R'$ and adjusted their treatment of St accordingly. Given the forcing implicit in the shearing vertical sheets of VSI \citep{Nelson_etal_2013, Stoll_Kley_2014}, a similar situation there seems inevitable. Other turbulent instabilities should be considered from this standpoint, as well. 

\section{Directly calculating $\alphat$, $\alpha$, and tangential stresses $\tanstress$ from driven turbulence simulations}\label{sec:alpha_alphat}

Driven turbulence calculations are quantified by an effective ``driving" $\alpha$ parameter.  Following the prescription described in \citet{Gole_etal_2020}, we write this control parameter as $\tilde\alpha$ 
\footnote{We note that  \citet{Gole_etal_2020} write $\tilde\alpha$ as $\alpha_{{\rm drive}}$.}
and it is defined as
\begin{equation}
    \tilde\alpha \equiv 
         \frac{1}{c^2}\sum_{i=x,y,z}\left(\overline{\langle |\delta v_i(x,y,z,t)|\rangle}\right)^2.
\end{equation}

where $c=1$ is the sound speed and $\delta v_i(x,y,z,t) = v_i(x,y,z,t) - \langle v_i\rangle_{xy}(z,t)$ is the velocity fluctuations. Here $\langle .. \rangle$ denotes the spatial average and the overbar denotes the temporal average with the $xy$ subscript being the spatial average over the $xy$ plane. For the temporal average in Eq. (\ref{eqn:alpha_sengupta}) we take the spatial averages at $5$ different snapshots at the statistically steady turbulent state.

Based on the foregoing discussion it follows that we can define and calculate the diffusive $\alpha$ in the simulation box by reference to the driving parameter $\tilde \alpha$ according to: 
\begin{equation}\label{eqn:alpha_sengupta}
    \alpha = \frac{1}{R'c^2}\sum_{i=x,y,z}\left(\overline{\langle |\delta v_i(x,y,z,t)|\rangle}\right)^2 = \frac{\Tilde{\alpha}}{R'}
\end{equation}
The Rossby number $R'$ can also be determined from information read directly from the KE power spectrum, according to the prescription described in Appendix \ref{sec:mixing_length}.

In the spirit of the above, we similarly express the spatio-temporal average of the tangential stresses $\mathtt{W}_{r\phi}$ in terms of its negative \textit{specific value} via the relationship $-\rho\tanstress$ where
\begin{equation}
    \tanstress
    \equiv 
    -\overline{\langle |\delta v_x(x,y,z,t)\delta v_y(x,y,z,t)|\rangle},
\end{equation}
where for practical intents and purposes $\rho$ is assumed to be constant.
Expressing $\mathtt{W}_{r\phi}$ in terms of negative quantity ensures that gas accretion in the traditional sense of the effect occurs toward the star.

\section{Mixing Length Phenomenology}\label{sec:mixing_length}

In light of our introductory remarks in Section 2.1, we seek to relate turbulent viscosity $D$ and $\alpha$ to the KE spectrum of the gas.
Diffusion in a turbulent medium can be phenomenologically assessed by the dominant length ($L$) and velocity scales ($V_L$) associated with the turbulent dynamics, which usually occurs at the start of the inertial range in 3D isotropic turbulence. According to standard mixing length theory
\citep[e.g.,][]{Knobloch_1978}, when the turbulence is isotropic, the diffusion becomes a scalar and can be written as 
\begin{equation}\label{eqn:mixingD}
    D=\frac{1}{3}V_LL.
\end{equation}
 Considering that a particular length scale $\ell$ reasonably contains two counter-rotating eddies of size $\ell/2$, equation \ref{eqn:mixingD} reads,
\begin{equation}\label{eqn:Dt}
    D=\frac{1}{3}V_LL = \frac{\pi}{3} \frac{V_L}{k_f} \sim \frac{V_L}{k_f}.
\end{equation}
Using Eqs. (\ref{eqn:velocity},\ref{eqn:omegaL}), Eq. (\ref{eqn:Dt}) above can be further manipulated into
\begin{equation}
    D = \frac{\sqrt{2k_fE(k_f)}}{k_f}
    =\frac{2k_fE(k_f)}{\sqrt{2k_f^3E(k_f)}}
    = \sqrt{\frac{2E(k_f)}{k_f}}.
    \label{Dt_spectral_rewrite}
\end{equation}

The approach leading to Eq. (\ref{Dt_spectral_rewrite}) captures both the form of the original mixing length ansatz and the Reynolds stress approach (Section \ref{sec:omegaL=omegaK}) as well as the spectral approach to identifying $V_L$, for which we generally find that indeed $V_L^2 \approx V_{\rm rms}^2 \sim \tilde\alpha c^2$.

{Using the expression for $\Omega_L$ found in Eq.(\ref{eqn:omegaL}), and $\tilde\alpha = V_L^2/c^2 = 2k_fE(k_f)/c^2 $, we write $R'$ in terms of $\tilde\alpha$ and $H\equiv c/\Omega_K$ as} 
\begin{equation}
  R' =  \Omega_L/\Omega_K = k_f\sqrt{2k_fE(k_f)}/\Omega_K = H k_f \sqrt{\tilde \alpha},
\end{equation}
which in turn leads to to an equivalent restatement of Eq. (\ref{Dt_spectral_rewrite}) as
\begin{equation}
    D_T = \frac{\alphat}{R'} H^2 \Omega_K = 
    \frac{\sqrt{\alphat}}{Hk_f} H^2 \Omega_K,
\end{equation}
which demonstrates the equivalence introduced in Section \ref{sec:spectra} between $\alpha$ and   $\tilde \alpha/R' = \sqrt{\alphat}/Hk_f $.
Thus in practice one can estimate $\alpha$ using either of these preceding formulae by identifying the wavenumber $k_f$ corresponding to the peak value of $v_k^2(k=k_f) = 2k_f E(k_f)$.

\section{ALGORITHM TO OBTAIN A PREDEFINED $\alpha$ and $R'$}\label{sec:algorithm}

An external forcing prescription is a simple way of generating the turbulence in the simulation which allows us to investigate particle dynamics and related processes, such as the formation of planetesimals under plausible nebula conditions. Below we present a simple algorithm for calibration of $R'$ and $\alpha$ in an externally forced turbulence in any numerical code one may use.
\begin{enumerate}
    \item Select a particular box size $(L_B = \xi H$, where $\xi<1$), and a forcing wavenumber $k_f$.
    \item Vary the energy injection by tuning the forcing parameter. In each case, generate a spectrum for turbulent energy \citet[See section 4.2.2 of][]{Sengupta_Umurhan_2023} and calculate $\Omega_L$ using equation \ref{eqn:omegaL} and $R'=\Omega_L/\Omega_K$. This will produce the calibration for $f_0$ vs $R'$. Note that for the correct values of $R'$, the wavenumbers must be scaled to the domain size such that $2\pi/k$ gives the actual physical length scale the wavenumber corresponds to.

    \item Calculate $\alpha$ for each case using equation \ref{eqn:alpha_sengupta} and the $R'$ calculated in step 2. The calibration for $f_0$ vs $\alpha$ is now done.
    \item Check for consistency, that $\alpha (R')$ follows a powerlaw relation with index $\sim 1$.
    \item For calibrations with a different $k_f$, repeat the above steps for the newly chosen wavenumber.
\end{enumerate}
 Once the calibration curves are in place, a value of $f_0$ can be easily chosen in order to achieve a desired value of $\alpha$. Note that in cases of particle-gas simulations, separate calibration curves need to be produced for each particle size. 

 Note that the $f_0$ vs $\alpha$ calibration presented in figure \ref{fig:figure1}b is applicable for gas-only simulations. When inertial (St$\neq 0$) particles are present, the interaction between the gas and the particles through  aerodynamic drag provides an extra source of non-linearity to the system, changing the overall dynamics. For example, in gas-only simulations, for $f_0\sim 0.03$ the expected value of $\Tilde{\alpha}$ is $\sim 3\times 10^{-3}$ (from calibrations in figure \ref{fig:figure1}a and b). When we introduce inertial particles with St$_K=0.06$ in order to investigate the particle scale heights (section \ref{sec:numericaltest}) we find $\Tilde{\alpha}\sim 1.6\times 10^{-3}$.  A reasonable rationalization to this would be to think of the increased effective density of the gas in the presence of particles and their mutual drag, making the largest eddy rotate slower with a smaller $\Omega_L$ and hence a smaller $R'$ for a fixed injection energy. Indeed \citet{Sengupta_Umurhan_2023} found this effect in their particle-gas simulations in a rotating stratified setup. However, at present, no proper phenomenology exists for rotating stratified turbulence \citep{Alexakis_Biferale_2010}, let alone in the presence of particles. So, the individual contributions from the rotation, stratification and particle drag cannot be identified separately and need a detailed investigation which we leave for  future work.

\section{Particle Scale Height}\label{sec:hpdetails}

For convenience we repeat here a simple derivation of the particle scale height, highlighting the different roles of ${\rm St}_K$ and ${\rm St}_L$. The reader is referred to the original papers for details \citep{ Cuzzi_etal_1993, Dubrulle_etal_1995, Cuzzi_Weidenschilling_2006, Youdin_Lithwick_2007, Zsom_2008, Sengupta_etal_2019}. The physics is to assume a steady state particle layer of some half-thickness $h$, in balance between downward mass flow per unit area $\dot{\sigma}_-$  at terminal settling velocity $V_T$
\begin{equation}
\dot{\sigma}_- = \rho_p V_T
\end{equation}
and upward gradient diffusion mass flow with particle diffusion coefficient $D_p$
\begin{equation}
\dot{\sigma}_+ = D_p (\partial \rho_p / \partial z).
\end{equation}
The settling velocity, for small particles with $\rm{St} \ll 1$, is simply $V_T = g t_s = {\rm St}_K h \Omega_K$ where $t_s$ is the particle stopping time and $g=\Omega_K^2 z$ is the vertical component of solar gravity at height $z$. The particle diffusion coefficient $D_p = D_{\rm gas}/(1+St_L^2) = \alpha c H/(1+St_L^2)$ \citep{Youdin_Lithwick_2007}. Note that the settling velocity depends on $St_K$ and the diffusion coefficient on $St_L$. One approximates $\rho_p \sim \rho_{po}$ and $(\partial \rho_p / \partial z) \sim \rho_{po}/h$, so 
\begin{eqnarray}
\rho_p V_T & = & D_p \rho_{po}/h \\
{\rm St}_K h \Omega_K & = & D_p/h \\
h^2 &=& \frac{\alpha c H}{{\rm St}_K \Omega_K (1+{\rm St}_L^2)}\\ 
h^2 &=& H^2 \frac{\alpha}{ {\rm St}_K (1+{\rm St}_L^2)}.
\end{eqnarray}
The ${\rm St}_L^2$ term from the diffusion coefficient plays relatively little role in the process, since ${\rm St}_L \ll 1$ for most realistic cases \citep{Carballido_etal_2011, Estrada_etal_2016}. This is also true in cases where there is a balance between radial diffusion and drift \citep{Dullemond_etal_2018, Umurhan_etal_2020} because radial drift, like vertical drift, is determined by ${\rm St}_K$.

\section{The Forcing Scheme}\label{sec:forcing}

In order to drive turbulence in the simulation box by external forcing we shall use the same prescription as \citet{Brandenburg_2001, Haugen_Brandenburg_2004} and \citet{Sengupta_Umurhan_2023}, and drive turbulence with a random, non-helical and delta-correlated forcing function $\bm{f}(\bm{x},t)$, written as
\begin{equation}
    \bm{f}(\bm{x},t) \equiv Re\left\{\mathcal{N}\bm{f}_{\bm{k}(t)}exp\left[i\bm{k}(t)\cdot \bm{x}+i\phi(t)\right]\right\}.
\end{equation}
Here $\bm{k}(t) = (k_x, k_y, k_z)$ is a time-dependent wavevector and $\phi(t)$ with $\left|\phi\right| < \pi$ is the random phase. On a purely dimensional argument, the normalization factor $\mathcal{N}$ can be written as $\mathcal{N}=f_0c(kc/\delta t)^{1/2}$, where $c$ is the sound speed set as $1$ in the code unit, $\delta t$ is the time step, $k=|\bm{k}|$ is the wavenumber and $f_0$ is a dimensionless factor we call the forcing amplitude. In the code, $f_0$ is the knob we adjust in order to control the forcing and achieve a desired value of $\alpha$ (Also, see below). We choose to force the system at some $k=k_f$, in which case, at each step a randomly chosen possible wavenumber within a narrow band of $k_f -0.5 < |\bm{k}| <k_f + 0.5$ is forced. The forcing is executed with the eigenfunctions of the curl operator 
\begin{equation}\label{eqn:forcing_term}
    \bm{f}_{\bm{k}}=\frac{i\bm{k}\times (\bm{k}\times \hat{e})-\sigma |\bm{k}|(\bm{k}\times \hat{e})}{\sqrt{1+\sigma^2}\bm{k}^2\sqrt{1-(\bm{k}\cdot\hat{e})^2/\bm{k}^2}}.
\end{equation}
Here $\hat{e}$ is the arbitrary unit vector used to generate $\bm{k}\times \hat{e}$ which is perpendicular to $\bm{k}$. $\sigma$ denotes the helicity factor which is set to zero in order to make the forcing purely non-helical. It is worth noting that this forcing is essentially divergenceless. However, as the fluid equations solved by the Pencil code are not strictly incompressible, which is perhaps more applicable for astrophysical systems compared to a fully incompressible dynamics , a small non-zero divergence is introduced over the course of the simulation.  Nonetheless, the spatio-temporal dynamics of all of our simulations are effectively incompressible \citep{Sengupta_Umurhan_2023}.

\section{Comments on Simulation Domain}

All variations of $(k_f, f_o)$ in the simulations shown here result in two lines on the plot of $(R', \alpha)$, one for each $k_f$. This is because all our simulations assumed the same computational box size $L_B = 0.2H$. It is simple to access other parts of $(R', \alpha)$ space, as follows. From equation \ref{eqn:length}, $L=H\sqrt{\alpha/R'}$. By definition, the integral scale $L$ can also be written as $L = L_B/k_f = \xi H / k_f$, where in the current simulations $\xi=0.2$. Setting these equal leads directly to $R'/\alpha = (k_f/\xi)^2$. This relationship can be used to move around in $(R', \alpha)$ space. Choosing a smaller $\xi$ allows us to reach smaller $\alpha$ for a given $R'$, or to decrease $k_f$ and capture a wider inertial range at some given $\alpha$. Conversely, choosing a larger $\xi$ allows us to reach larger $\alpha$ for a given $R'$. The details of these results will vary however, from code to code.  Moreover, the scaling is not entirely linear unless the forced turbulence is homogeneous and isotropic; increasing $\xi$ too much in the presence of nebula rotation, stratification and shear can lead to a more complicated blend of rotation and stratification.

\section{Inferring S\lowercase{t}$_{\rm L}$ and $R'$ from Observations}\label{sec:revised_alphaSt_from_obs}

Here we sketch how combining various independently observed properties of a turbulent protoplanetary disk can lead to constraints not only on its true turbulent $\alpha$, but on both St$_{\rm L}$ which is of value in its own right and also the turbulent $R'$, which may be diagnostic of the {\it kind} of turbulent process that is active. Even if observational realities preclude such a clean extraction, being aware of the implications of $R' > 1$ is important. 
The (potentially) observed quantities are the particle size $r$ and actual internal density $\rho_p$ (the latter admitting the possibility the particles are porous aggregates), the disk gas surface mass density $\Sigma_g$, the local rotation rate and temperature, the root-mean-square gas turbulent velocity ($V_{\text{rms}}$) through spectral line broadening (Eq. \ref{def:alphat_sims}), and 
the particle vertical scale height $h$ from mm-cm interferometric observations, usually expressed as a ratio to the local gas scale height $H$.

Eq. (\ref{eqn:scaleheight}), which is valid in the limit $St \gg \alpha$, is directly rewritten as
\begin{equation}
\frac{h^2}{H^2} = 
\frac{\tilde \alpha}{R'{\rm St}_K(1+{R'}^2{\rm St}_K^2)},
\label{eqn:Rprime_equation_obs}
\end{equation}
where we have used the relationships
$\alpha\equiv \tilde\alpha/R'$ found in Eq. (\ref{eqn:alpha_alphatilde}) and the definition ${\rm St}_L \equiv 
R'{\rm St}_K$. 
Eq. (\ref{eqn:Rprime_equation_obs}) can be viewed as a cubic equation for ${\rm St}_L$, in which the turbulent Stokes number is a function of the observables $h/H$ and $\tilde\alpha \sim V_{{\rm rms}}^2/c^2$ (see Appendix \ref{sec:revised_alphaSt_from_obs} for details):
\begin{equation}
{\rm St}_L^3 + {\rm St}_L = \varphi,
\label{eqn:cubic_StL}
\end{equation}
where 
\begin{equation}
\varphi \equiv \frac{H^2 \tilde \alpha}{h^2}
\approx 
\frac{H^2 \tilde V^2_{{\rm rms}}}{h^2 c^2} = 
\frac{\tilde V^2_{{\rm rms}}}{h^2 \Omega_K^2}
.
\end{equation}
Cubic equations are awkward to solve but one can obtain simple expressions in two limiting regimes to illustrate the use of the technique. It is important to note that $\St_L$ {\it itself} may be determined directly from $\varphi$, which is composed entirely of the observable quantities, $h,V_{{\rm rms}}$, and $\Omega_K$.

First consider the highly likely regime $\varphi \ll 1$ or $\St_L =R'\St_K \ll 1$. In this regime $\St_L \gg \St_L^3$, so 
from equation \ref{eqn:cubic_StL}, $R'\St_K = \St_L \approx \varphi = \alphat(H/h)^2$. We can now solve directly for $R'$ in terms of observables, making  use of the standard approximation for the small-particle (Epstein) regime that ${\rm St}_K= t_s \Omega_K = 2 r \rho_p /\Sigma_g$. 
\begin{equation}
R'=\frac{\alphat}{{\rm St}_K}\left(\frac{H}{h} \right)^2   = \frac{\Sigma_g}{2 r \rho_p}  \alphat \left(\frac{H}{h} \right)^2  ;
\end{equation}
that is, $R'$ is constrained by a combination of potentially observable parameters including $\alphat$, and once $R'$ is determined, $\alpha$ itself can be immediately calculated.

The alternate limiting regime is $\varphi \gg 1$ or $\St_L =R'\St_K \gg 1$. Here, $\St^3_L \gg \St_L$, so 
from equation \ref{eqn:cubic_StL}, $R'\St_K = \St_L \approx \varphi^{1/3}$ and
\begin{equation}
R'= \frac{\alphat^{1/3}}{{\rm St}_K}\left(\frac{H}{h} \right)^{2/3}  = \frac{\Sigma_g }{2 r \rho_p} \alphat^{1/3} \left(\frac{H}{h} \right)^{2/3}.
\end{equation}
It will always be evident from the observed values of $h/H$ (ALMA or VLA) and $\alphat$ (spectral line broadening) which regime is of interest. Again, once $R'$ is determined, $\alpha$ itself can be immediately calculated. The usually cited combined parameter loosely referred to as $\alpha/{\rm St}$ is thus seen to be merely an intermediate step in a complete determination of the properties of the system.

\bibliographystyle{aasjournal}
\bibliography{reference}

\begin{thebibliography}{}
\expandafter\ifx\csname natexlab\endcsname\relax\def\natexlab#1{#1}\fi
\providecommand{\url}[1]{\href{#1}{#1}}
\providecommand{\dodoi}[1]{doi:~\href{http://doi.org/#1}{\nolinkurl{#1}}}
\providecommand{\doeprint}[1]{\href{http://ascl.net/#1}{\nolinkurl{http://ascl.net/#1}}}
\providecommand{\doarXiv}[1]{\href{https://arxiv.org/abs/#1}{\nolinkurl{https://arxiv.org/abs/#1}}}

\bibitem[{Alexakis \& Biferale(2018)}]{Alexakis_Biferale_2010}
Alexakis, A., \& Biferale, L. 2018, Physics Reports, 767-769, 1,
  \dodoi{https://doi.org/10.1016/j.physrep.2018.08.001}

\bibitem[{{Balbus} \& {Hawley}(1991)}]{balbus_Hawley_1991}
{Balbus}, S.~A., \& {Hawley}, J.~F. 1991, \apj, 376, 214,
  \dodoi{10.1086/170270}

\bibitem[{{Balbus} \& {Papaloizou}(1999)}]{Balbus_Papaloizou_1999}
{Balbus}, S.~A., \& {Papaloizou}, J. C.~B. 1999, \apj, 521, 650,
  \dodoi{10.1086/307594}

\bibitem[{{Birnstiel} {et~al.}(2010){Birnstiel}, {Dullemond}, \&
  {Brauer}}]{Birnstiel_etal_2010}
{Birnstiel}, T., {Dullemond}, C.~P., \& {Brauer}, F. 2010, \aap, 513, A79,
  \dodoi{10.1051/0004-6361/200913731}

\bibitem[{{Birnstiel} {et~al.}(2016){Birnstiel}, {Fang}, \&
  {Johansen}}]{Birnstiel_etal_2016}
{Birnstiel}, T., {Fang}, M., \& {Johansen}, A. 2016, \ssr, 205, 41,
  \dodoi{10.1007/s11214-016-0256-1}

\bibitem[{{Brandenburg}(2001)}]{Brandenburg_2001}
{Brandenburg}, A. 2001, \apj, 550, 824, \dodoi{10.1086/319783}

\bibitem[{{Cameron}(1978)}]{Cameron_1978}
{Cameron}, A.~G.~W. 1978, Moon and Planets, 18, 5, \dodoi{10.1007/BF00896696}

\bibitem[{{Carballido} {et~al.}(2011){Carballido}, {Bai}, \&
  {Cuzzi}}]{Carballido_etal_2011}
{Carballido}, A., {Bai}, X.-N., \& {Cuzzi}, J.~N. 2011, \mnras, 415, 93,
  \dodoi{10.1111/j.1365-2966.2011.18661.x}

\bibitem[{{Carrasco-Gonz{\'a}lez} {et~al.}(2019){Carrasco-Gonz{\'a}lez},
  {Sierra}, {Flock}, {Zhu}, {Henning}, {Chandler}, {Galv{\'a}n-Madrid},
  {Mac{\'\i}as}, {Anglada}, {Linz}, {Osorio}, {Rodr{\'\i}guez}, {Testi},
  {Torrelles}, {P{\'e}rez}, \& {Liu}}]{Carrascogonzalez_etal_2019}
{Carrasco-Gonz{\'a}lez}, C., {Sierra}, A., {Flock}, M., {et~al.} 2019, \apj,
  883, 71, \dodoi{10.3847/1538-4357/ab3d33}

\bibitem[{{Coleman} {et~al.}(1992){Coleman}, {Ferziger}, \&
  {Spalart}}]{Coleman_etal_1992}
{Coleman}, G.~N., {Ferziger}, J.~H., \& {Spalart}, P.~R. 1992, Journal of Fluid
  Mechanics, 244, 677, \dodoi{10.1017/S0022112092003264}

\bibitem[{Cushman-Roisin(2011)}]{Cushman-Roisin-Book}
Cushman-Roisin, B. 2011, Introduction to geophysical fluid dynamics : physical
  and numerical aspects, 2nd edn., International geophysics series ; v. 101
  (Waltham, MA: Academic Press)

\bibitem[{{Cuzzi} {et~al.}(1993){Cuzzi}, {Dobrovolskis}, \&
  {Champney}}]{Cuzzi_etal_1993}
{Cuzzi}, J.~N., {Dobrovolskis}, A.~R., \& {Champney}, J.~M. 1993, \icarus, 106,
  102, \dodoi{10.1006/icar.1993.1161}

\bibitem[{{Cuzzi} {et~al.}(2010){Cuzzi}, {Hogan}, \&
  {Bottke}}]{Cuzzi_etal_2010}
{Cuzzi}, J.~N., {Hogan}, R.~C., \& {Bottke}, W.~F. 2010, \icarus, 208, 518,
  \dodoi{10.1016/j.icarus.2010.03.005}

\bibitem[{{Cuzzi} {et~al.}(2001){Cuzzi}, {Hogan}, {Paque}, \&
  {Dobrovolskis}}]{Cuzzi_etal_2001}
{Cuzzi}, J.~N., {Hogan}, R.~C., {Paque}, J.~M., \& {Dobrovolskis}, A.~R. 2001,
  \apj, 546, 496, \dodoi{10.1086/318233}

\bibitem[{{Cuzzi} \& {Weidenschilling}(2006)}]{Cuzzi_Weidenschilling_2006}
{Cuzzi}, J.~N., \& {Weidenschilling}, S.~J. 2006, in Meteorites and the Early
  Solar System II, ed. D.~S. {Lauretta} \& H.~Y. {McSween}, 353

\bibitem[{{Davidson}(2004)}]{Davidson_2004}
{Davidson}, P.~A. 2004, {Turbulence : an introduction for scientists and
  engineers}

\bibitem[{{Doi} \& {Kataoka}(2021)}]{Doi_Kataoka_2021}
{Doi}, K., \& {Kataoka}, A. 2021, \apj, 912, 164,
  \dodoi{10.3847/1538-4357/abe5a6}

\bibitem[{{Doi} \& {Kataoka}(2023)}]{Doi_Kataoka_2023}
---. 2023, \apj, 957, 11, \dodoi{10.3847/1538-4357/acf5df}

\bibitem[{{Dubrulle}(1992)}]{Dubrulle_1992}
{Dubrulle}, B. 1992, \aap, 266, 592

\bibitem[{{Dubrulle} {et~al.}(1995){Dubrulle}, {Morfill}, \&
  {Sterzik}}]{Dubrulle_etal_1995}
{Dubrulle}, B., {Morfill}, G., \& {Sterzik}, M. 1995, \icarus, 114, 237,
  \dodoi{10.1006/icar.1995.1058}

\bibitem[{{Dullemond} {et~al.}(2018){Dullemond}, {Birnstiel}, {Huang},
  {Kurtovic}, {Andrews}, {Guzm{\'a}n}, {P{\'e}rez}, {Isella}, {Zhu}, {Benisty},
  {Wilner}, {Bai}, {Carpenter}, {Zhang}, \& {Ricci}}]{Dullemond_etal_2018}
{Dullemond}, C.~P., {Birnstiel}, T., {Huang}, J., {et~al.} 2018, \apjl, 869,
  L46, \dodoi{10.3847/2041-8213/aaf742}

\bibitem[{{Estrada} \& {Cuzzi}(2016)}]{Estrada_Cuzzi_2016}
{Estrada}, P.~R., \& {Cuzzi}, J.~N. 2016, in Lunar and Planetary Science
  Conference, 2854

\bibitem[{{Estrada} \& {Cuzzi}(2022)}]{Estrada_etal_2022}
{Estrada}, P.~R., \& {Cuzzi}, J.~N. 2022, \apj, in preparation

\bibitem[{{Estrada} {et~al.}(2016){Estrada}, {Cuzzi}, \&
  {Morgan}}]{Estrada_etal_2016}
{Estrada}, P.~R., {Cuzzi}, J.~N., \& {Morgan}, D.~A. 2016, \apj, 818, 200,
  \dodoi{10.3847/0004-637X/818/2/200}

\bibitem[{{Estrada} {et~al.}(2022){Estrada}, {Cuzzi}, \&
  {Umurhan}}]{Estrada_etal_2021}
{Estrada}, P.~R., {Cuzzi}, J.~N., \& {Umurhan}, O.~M. 2022, \apj, in
  preparation

\bibitem[{{Flaherty} {et~al.}(2020){Flaherty}, {Hughes}, {Simon}, {Qi}, {Bai},
  {Bulatek}, {Andrews}, {Wilner}, \& {K{\'o}sp{\'a}l}}]{Flaherty_etal_2020}
{Flaherty}, K., {Hughes}, A.~M., {Simon}, J.~B., {et~al.} 2020, \apj, 895, 109,
  \dodoi{10.3847/1538-4357/ab8cc5}

\bibitem[{{Flaherty} {et~al.}(2018){Flaherty}, {Hughes}, {Teague}, {Simon},
  {Andrews}, \& {Wilner}}]{Flaherty_etal_2018}
{Flaherty}, K.~M., {Hughes}, A.~M., {Teague}, R., {et~al.} 2018, \apj, 856,
  117, \dodoi{10.3847/1538-4357/aab615}

\bibitem[{{Flaherty} {et~al.}(2017){Flaherty}, {Hughes}, {Rose}, {Simon}, {Qi},
  {Andrews}, {K{\'o}sp{\'a}l}, {Wilner}, {Chiang}, {Armitage}, \&
  {Bai}}]{flaherty_etal_2017}
{Flaherty}, K.~M., {Hughes}, A.~M., {Rose}, S.~C., {et~al.} 2017, \apj, 843,
  150, \dodoi{10.3847/1538-4357/aa79f9}

\bibitem[{{Flock} {et~al.}(2020){Flock}, {Turner}, {Nelson}, {Lyra}, {Manger},
  \& {Klahr}}]{Flock_etal_2020}
{Flock}, M., {Turner}, N.~J., {Nelson}, R.~P., {et~al.} 2020, \apj, 897, 155,
  \dodoi{10.3847/1538-4357/ab9641}

\bibitem[{{Gole} {et~al.}(2020){Gole}, {Simon}, {Li}, {Youdin}, \&
  {Armitage}}]{Gole_etal_2020}
{Gole}, D.~A., {Simon}, J.~B., {Li}, R., {Youdin}, A.~N., \& {Armitage}, P.~J.
  2020, \apj, 904, 132, \dodoi{10.3847/1538-4357/abc334}

\bibitem[{{Hartlep} \& {Cuzzi}(2020)}]{Hartlep_Cuzzi_2020}
{Hartlep}, T., \& {Cuzzi}, J.~N. 2020, \apj, 892, 120,
  \dodoi{10.3847/1538-4357/ab76c3}

\bibitem[{{Hartlep} {et~al.}(2017){Hartlep}, {Cuzzi}, \&
  {Weston}}]{Hartlep_etal_2017}
{Hartlep}, T., {Cuzzi}, J.~N., \& {Weston}, B. 2017, \pre, 95, 033115,
  \dodoi{10.1103/PhysRevE.95.033115}

\bibitem[{{Haugen} \& {Brandenburg}(2004)}]{Haugen_Brandenburg_2004}
{Haugen}, N. E.~L., \& {Brandenburg}, A. 2004, \pre, 70, 026405,
  \dodoi{10.1103/PhysRevE.70.026405}

\bibitem[{{Hughes} \& {Armitage}(2010)}]{Hughes_Armitage_2010}
{Hughes}, A. L.~H., \& {Armitage}, P.~J. 2010, \apj, 719, 1633,
  \dodoi{10.1088/0004-637X/719/2/1633}

\bibitem[{{Hughes} \& {Armitage}(2012)}]{Hughes_Armitage_2012}
---. 2012, \mnras, 423, 389, \dodoi{10.1111/j.1365-2966.2012.20892.x}

\bibitem[{{Jacquet} {et~al.}(2012){Jacquet}, {Gounelle}, \&
  {Fromang}}]{Jacquet_etal_2012}
{Jacquet}, E., {Gounelle}, M., \& {Fromang}, S. 2012, \icarus, 220, 162,
  \dodoi{10.1016/j.icarus.2012.04.022}

\bibitem[{{Johansen} {et~al.}(2006){Johansen}, {Klahr}, \&
  {Mee}}]{Johansen_et_al_2006}
{Johansen}, A., {Klahr}, H., \& {Mee}, A.~J. 2006, \mnras, 370, L71,
  \dodoi{10.1111/j.1745-3933.2006.00191.x}

\bibitem[{{Johansen} {et~al.}(2007){Johansen}, {Oishi}, {Mac Low}, {Klahr},
  {Henning}, \& {Youdin}}]{Johansen_etal_2007}
{Johansen}, A., {Oishi}, J.~S., {Mac Low}, M.-M., {et~al.} 2007, \nat, 448,
  1022, \dodoi{10.1038/nature06086}

\bibitem[{{Knobloch}(1978)}]{Knobloch_1978}
{Knobloch}, E. 1978, \apj, 225, 1050, \dodoi{10.1086/156572}

\bibitem[{{Lesur} {et~al.}(2022){Lesur}, {Ercolano}, {Flock}, {Lin}, {Yang},
  {Barranco}, {Benitez-Llambay}, {Goodman}, {Johansen}, {Klahr}, {Laibe},
  {Lyra}, {Marcus}, {Nelson}, {Squire}, {Simon}, {Turner}, {Umurhan}, \&
  {Youdin}}]{Lesur_etal_2022}
{Lesur}, G., {Ercolano}, B., {Flock}, M., {et~al.} 2022, arXiv e-prints,
  arXiv:2203.09821.
\newblock \doarXiv{2203.09821}

\bibitem[{{Lin} \& {Bodenheimer}(1982)}]{Lin_Bodenheimer_1982}
{Lin}, D.~N.~C., \& {Bodenheimer}, P. 1982, \apj, 262, 768,
  \dodoi{10.1086/160472}

\bibitem[{{Lynden-Bell} \& {Pringle}(1974)}]{LyndenBell_Pringle_1974}
{Lynden-Bell}, D., \& {Pringle}, J.~E. 1974, \mnras, 168, 603,
  \dodoi{10.1093/mnras/168.3.603}

\bibitem[{{Lyra}(2014)}]{Lyra_2014}
{Lyra}, W. 2014, \apj, 789, 77, \dodoi{10.1088/0004-637X/789/1/77}

\bibitem[{{Mac{\'\i}as} {et~al.}(2021){Mac{\'\i}as}, {Guerra-Alvarado},
  {Carrasco-Gonz{\'a}lez}, {Ribas}, {Espaillat}, {Huang}, \&
  {Andrews}}]{Macias_etal_2021}
{Mac{\'\i}as}, E., {Guerra-Alvarado}, O., {Carrasco-Gonz{\'a}lez}, C., {et~al.}
  2021, \aap, 648, A33, \dodoi{10.1051/0004-6361/202039812}

\bibitem[{{Marcus} {et~al.}(2013){Marcus}, {Pei}, {Jiang}, \&
  {Hassanzadeh}}]{Marcus_etal_2013}
{Marcus}, P.~S., {Pei}, S., {Jiang}, C.-H., \& {Hassanzadeh}, P. 2013, \prl,
  111, 084501, \dodoi{10.1103/PhysRevLett.111.084501}

\bibitem[{{Morfill}(1985)}]{Morfill_1985}
{Morfill}, G.~E. 1985, in Birth and Infancy of Stars, 693

\bibitem[{{Nelson} {et~al.}(2013){Nelson}, {Gressel}, \&
  {Umurhan}}]{Nelson_etal_2013}
{Nelson}, R.~P., {Gressel}, O., \& {Umurhan}, O.~M. 2013, \mnras, 435, 2610,
  \dodoi{10.1093/mnras/stt1475}

\bibitem[{{Ormel} \& {Cuzzi}(2007)}]{Ormel_Cuzzi_2007}
{Ormel}, C.~W., \& {Cuzzi}, J.~N. 2007, \aap, 466, 413,
  \dodoi{10.1051/0004-6361:20066899}

\bibitem[{{Pencil Code Collaboration} {et~al.}(2021){Pencil Code
  Collaboration}, {Brandenburg}, {Johansen}, {Bourdin}, {Dobler}, {Lyra},
  {Rheinhardt}, {Bingert}, {Haugen}, {Mee}, {Gent}, {Babkovskaia}, {Yang},
  {Heinemann}, {Dintrans}, {Mitra}, {Candelaresi}, {Warnecke},
  {K{\"a}pyl{\"a}}, {Schreiber}, {Chatterjee}, {K{\"a}pyl{\"a}}, {Li},
  {Kr{\"u}ger}, {Aarnes}, {Sarson}, {Oishi}, {Schober}, {Plasson}, {Sandin},
  {Karchniwy}, {Rodrigues}, {Hubbard}, {Guerrero}, {Snodin}, {Losada},
  {Pekkil{\"a}}, \& {Qian}}]{Pencil_code}
{Pencil Code Collaboration}, {Brandenburg}, A., {Johansen}, A., {et~al.} 2021,
  The Journal of Open Source Software, 6, 2807, \dodoi{10.21105/joss.02807}

\bibitem[{{Pizzati} {et~al.}(2023){Pizzati}, {Rosotti}, \&
  {Tabone}}]{Pizzati_etal_2023}
{Pizzati}, E., {Rosotti}, G.~P., \& {Tabone}, B. 2023, \mnras, 524, 3184,
  \dodoi{10.1093/mnras/stad2057}

\bibitem[{{Prendergast} \& {Burbidge}(1968)}]{Prendergast_Burbidge_1968}
{Prendergast}, K.~H., \& {Burbidge}, G.~R. 1968, \apjl, 151, L83,
  \dodoi{10.1086/180148}

\bibitem[{{Safronov}(1972)}]{Safronov_1972}
{Safronov}, V.~S. 1972, {Evolution of the protoplanetary cloud and formation of
  the earth and planets.}

\bibitem[{{Sekiya}(1998)}]{Sekiya_1998}
{Sekiya}, M. 1998, \icarus, 133, 298, \dodoi{10.1006/icar.1998.5933}

\bibitem[{{Sengupta} {et~al.}(2019){Sengupta}, {Dodson-Robinson}, {Hasegawa},
  \& {Turner}}]{Sengupta_etal_2019}
{Sengupta}, D., {Dodson-Robinson}, S.~E., {Hasegawa}, Y., \& {Turner}, N.~J.
  2019, \apj, 874, 26, \dodoi{10.3847/1538-4357/aafc36}

\bibitem[{{Sengupta} {et~al.}(2022){Sengupta}, {Estrada}, {Cuzzi}, \&
  {Humayun}}]{Sengupta_etal_2022}
{Sengupta}, D., {Estrada}, P.~R., {Cuzzi}, J.~N., \& {Humayun}, M. 2022, \apj,
  932, 82, \dodoi{10.3847/1538-4357/ac6dcc}

\bibitem[{{Sengupta} \& {Umurhan}(2023)}]{Sengupta_Umurhan_2023}
{Sengupta}, D., \& {Umurhan}, O.~M. 2023, \apj, 942, 74,
  \dodoi{10.3847/1538-4357/ac9411}

\bibitem[{{Shakura} \& {Sunyaev}(1973)}]{Shakura_Sunyaev_1973}
{Shakura}, N.~I., \& {Sunyaev}, R.~A. 1973, \aap, 24, 337

\bibitem[{{Shakura} {et~al.}(1978){Shakura}, {Sunyaev}, \&
  {Zilitinkevich}}]{Shakura_etal_1978}
{Shakura}, N.~I., {Sunyaev}, R.~A., \& {Zilitinkevich}, S.~S. 1978, \aap, 62,
  179

\bibitem[{{Stamper} \& {Taylor}(2017)}]{Stamper_Taylor_2017}
{Stamper}, M.~A., \& {Taylor}, J.~R. 2017, Ocean Dynamics, 67, 65,
  \dodoi{10.1007/s10236-016-1011-6}

\bibitem[{{Stoll} \& {Kley}(2014)}]{Stoll_Kley_2014}
{Stoll}, M. H.~R., \& {Kley}, W. 2014, \aap, 572, A77,
  \dodoi{10.1051/0004-6361/201424114}

\bibitem[{{Stoll} \& {Kley}(2016)}]{Stoll_Kley_2016}
---. 2016, \aap, 594, A57, \dodoi{10.1051/0004-6361/201527716}

\bibitem[{{Teague} {et~al.}(2018){Teague}, {Henning}, {Guilloteau}, {Bergin},
  {Semenov}, {Dutrey}, {Flock}, {Gorti}, \& {Birnstiel}}]{Teague_etal_2018}
{Teague}, R., {Henning}, T., {Guilloteau}, S., {et~al.} 2018, \apj, 864, 133,
  \dodoi{10.3847/1538-4357/aad80e}

\bibitem[{{Tennekes} \& {Lumley}(1972)}]{Tennekes_Lumley_1972}
{Tennekes}, H., \& {Lumley}, J.~L. 1972, {First Course in Turbulence}

\bibitem[{{Turner} {et~al.}(2014){Turner}, {Fromang}, {Gammie}, {Klahr},
  {Lesur}, {Wardle}, \& {Bai}}]{Turner_etal_2014}
{Turner}, N.~J., {Fromang}, S., {Gammie}, C., {et~al.} 2014, in Protostars and
  Planets VI, ed. H.~{Beuther}, R.~S. {Klessen}, C.~P. {Dullemond}, \&
  T.~{Henning}, 411, \dodoi{10.2458/azu_uapress_9780816531240-ch018}

\bibitem[{Umurhan {et~al.}(2020)Umurhan, Estrada, \& Cuzzi}]{Umurhan_etal_2020}
Umurhan, O.~M., Estrada, P.~R., \& Cuzzi, J.~N. 2020, \apj, 895, 26,
  \dodoi{10.3847/1538-4357/ab899d}

\bibitem[{{Villenave} {et~al.}(2022){Villenave}, {Stapelfeldt}, {Duch{\^e}ne},
  {M{\'e}nard}, {Lambrechts}, {Sierra}, {Flores}, {Dent}, {Wolff}, {Ribas},
  {Benisty}, {Cuello}, \& {Pinte}}]{Villenave_etal_2022}
{Villenave}, M., {Stapelfeldt}, K.~R., {Duch{\^e}ne}, G., {et~al.} 2022, \apj,
  930, 11, \dodoi{10.3847/1538-4357/ac5fae}

\bibitem[{{Villenave} {et~al.}(2023){Villenave}, {Podio}, {Duch{\^e}ne},
  {Stapelfeldt}, {Melis}, {Carrasco-Gonzalez}, {Le Gouellec}, {M{\'e}nard}, {de
  Simone}, {Chandler}, {Garufi}, {Pinte}, {Bianchi}, \&
  {Codella}}]{Villenave_etal_2023}
{Villenave}, M., {Podio}, L., {Duch{\^e}ne}, G., {et~al.} 2023, \apj, 946, 70,
  \dodoi{10.3847/1538-4357/acb92e}

\bibitem[{{V\"{o}lk} {et~al.}(1980){V\"{o}lk}, {Jones}, {Morfill}, \&
  {Roeser}}]{Voelk_etal_1980}
{V\"{o}lk}, H.~J., {Jones}, F.~C., {Morfill}, G.~E., \& {Roeser}, S. 1980,
  \aap, 85, 316

\bibitem[{{Weidenschilling}(1984)}]{Weidenschilling_1984}
{Weidenschilling}, S.~J. 1984, \icarus, 60, 553,
  \dodoi{10.1016/0019-1035(84)90164-7}

\bibitem[{{Weidenschilling}(1997)}]{Weidenschilling_1997}
---. 1997, \icarus, 127, 290, \dodoi{10.1006/icar.1997.5712}

\bibitem[{{Weidenschilling} \& {Cuzzi}(1993)}]{Weidenschilling_Cuzzi_1993}
{Weidenschilling}, S.~J., \& {Cuzzi}, J.~N. 1993, in Protostars and Planets
  III, ed. E.~H. {Levy} \& J.~I. {Lunine}, 1031

\bibitem[{{Youdin} \& {Goodman}(2005)}]{Youdin_Goodman_2005}
{Youdin}, A.~N., \& {Goodman}, J. 2005, \apj, 620, 459, \dodoi{10.1086/426895}

\bibitem[{{Youdin} \& {Lithwick}(2007)}]{Youdin_Lithwick_2007}
{Youdin}, A.~N., \& {Lithwick}, Y. 2007, \icarus, 192, 588,
  \dodoi{10.1016/j.icarus.2007.07.012}

\bibitem[{{Zahn}(1989)}]{Zahn_1989}
{Zahn}, J.-P. 1989, {Turbulent transport in Stellar Rotation Zones: Causes and
  Effects}, 183

\bibitem[{{Zsom} \& {Dullemond}(2008)}]{Zsom_2008}
{Zsom}, A., \& {Dullemond}, C.~P. 2008, \aap, 489, 931,
  \dodoi{10.1051/0004-6361:200809921}

\bibitem[{{Zsom} {et~al.}(2010){Zsom}, {Ormel}, {G{\"u}ttler}, {Blum}, \&
  {Dullemond}}]{Zsom_etal_2010}
{Zsom}, A., {Ormel}, C.~W., {G{\"u}ttler}, C., {Blum}, J., \& {Dullemond},
  C.~P. 2010, \aap, 513, A57, \dodoi{10.1051/0004-6361/200912976}

\end{thebibliography}

\end{document}